\documentclass[aps,prd,onecolumn,showpacs,showkeys,superscriptaddress]{revtex4-2}
\usepackage{latexsym}
\usepackage{amssymb}
\usepackage{amsmath}
\usepackage{amscd}
\usepackage{amsthm}
\usepackage{graphicx}
\usepackage{textcomp}
\usepackage{colortbl}
\usepackage[colorlinks]{hyperref}
\usepackage{hyperref}
\usepackage[font={footnotesize,it}]{caption}
\usepackage{multirow}
\usepackage[T1]{fontenc}
\usepackage{ae,aecompl}
\usepackage{subcaption}
\RequirePackage{graphicx}
\RequirePackage{float}
\RequirePackage{hyperref}
\RequirePackage{amsmath}
\RequirePackage{amssymb}
\RequirePackage{mathtools}
\usepackage{color}
\usepackage{comment}
\usepackage{xcolor}
\usepackage{booktabs} 

\begin{document}
\title{Dark Energy Phenomenology in a $f(R,\Sigma,T)$ Gravity Framework: $O_m(z)$ Parameterization Approach}
\vspace{10mm}
\author{ N. Myrzakulov}
\email{nmyrzakulov@gmail.com}
\affiliation{L. N. Gumilyov Eurasian National University, Astana 010008, Kazakhstan}

\author{Anirudh Pradhan
}
\email{pradhan.anirudh@gmail.com}
\affiliation{Centre for Cosmology, Astrophysics and Space Science (CCASS), GLA University, Mathura-281406, U.P., India.}

\author{Anil Kumar Yadav}
\email[]{abanilyadav@yahoo.co.in}
\affiliation{Department of Physics, United College of Engineering and Research,Greater Noida - 201310, India}

\author{S. H. Shekh}
\email[]{da\_salim@rediff.com}
\affiliation{Department of Mathematics, S.P.M. Science and Gilani Arts, Commerce College, Ghatanji, Yavatmal, \\Maharashtra-445301, India}
\affiliation{L N Gumilyov Eurasian National University, Astana 010008, Kazakhstan}
\affiliation{Pacif Institute of Cosmology and Selfology (PICS), Sagara, Sambalpur, Odisha 768224, India}

\vspace{10mm}
\begin{abstract} \textbf{Abstract:}
This study investigates the cosmological implications of $f(R,\Sigma,T)$ gravity by reconstructing the Hubble parameter from a logarithmic parameterization of the $O_m(z)$ diagnostic. We derive the field equations for $f(R,\Sigma,T) = R + \Sigma + 2\pi\eta T$ within the homogeneous, isotropic, and spatially flat Friedmann-Robertson-Walker (FRW) metric. A comprehensive analysis of key physical parameters, including the equation of state (EoS) parameter $(\omega)$, the $(\omega-\omega')$-plane, the squared sound speed $\vartheta^2$, and various energy conditions (Null, Dominant, Strong), is presented. Our findings reveal a dynamic EoS parameter that consistently remains within the quintessence regime ($-1 < \omega < -1/3$), approaching $\omega = -1$ in the far future, thereby avoiding phantom behavior and maintaining the weak energy condition. The model successfully reproduces the cosmic transition from deceleration to acceleration, as indicated by the deceleration parameter $q(z)$ crossing zero. While the model aligns well with observational data for cosmic expansion, the analysis of $\vartheta^2$ indicates classical instability, a point requiring further theoretical refinement. Overall, this work demonstrates the viability of $f(R,\Sigma,T)$ gravity as a framework capable of describing the universe's accelerated expansion, consistent with current cosmological observations.\\
	

\end{abstract}
\maketitle
\section{Introduction}

The discovery of the accelerated expansion of the Universe at the end of the 20th century marked a revolutionary breakthrough in modern cosmology. The initial evidence for this unexpected cosmic behavior was provided by the independent observations of distant Type Ia supernovae (SNe Ia) by the Supernova Cosmology Project and the High-z Supernova Search Team \cite{Perlmutter1999, Riess1998}. Since then, a wealth of complementary observations, including the Cosmic Microwave Background (CMB) anisotropies \cite{Planck2020}, Baryon Acoustic Oscillations (BAO) and large-scale structure surveys, have consistently supported the existence of an accelerated phase in the cosmic expansion history. This late-time cosmic acceleration challenges the conventional understanding of gravity described by General Relativity (GR), as the standard Einstein equations with ordinary matter and radiation components predict a decelerating universe.

To account for this accelerating behavior, cosmologists have postulated the existence of a mysterious component termed dark energy (DE), which contributes roughly 70\% of the total energy density of the Universe \cite{Planck2020}. The simplest and most widely accepted candidate for dark energy is the cosmological constant ($\Lambda$), introduced originally by Einstein, corresponding to a constant vacuum energy density with an equation of state parameter $\omega = -1$. Despite the success of the $\Lambda$ Cold Dark Matter ($\Lambda$CDM) model in explaining a broad range of cosmological observations \cite{Planck2020, Scolnic2018}, it suffers from significant theoretical challenges, such as the fine-tuning problem and the coincidence problem \cite{Weinberg1989, Sahni2000}. These issues have motivated the exploration of alternative explanations, including both dynamic dark energy models and modifications to the theory of gravity itself. In this context, modified theories of gravity have emerged as promising frameworks to explain the accelerated expansion without invoking exotic forms of matter-energy content.

Modified theories of gravity extend or generalize Einstein's GR by altering the gravitational sector of the field equations. A wide variety of such modifications have been proposed over the past two decades, each introducing different corrections to the standard Einstein-Hilbert action. The most studied modified theories are $f(R)$, $f(T)$, Gauss-Bonnet gravity, rane-world models \cite{Sotiriou2010, DeFelice2010,Cai2016,Jimenez2018,Fujii2003,Nojiri2005,Dvali2000,Nojiri2011,Capozziello2011,Jimenez2018}.  All these approaches represent attempts to generalize gravity's geometric foundation while still reproducing observational successes of GR at solar system scales. The $f(R,T)$ gravity theory proposed by Harko et al. \cite{Harko2011,Moraes2016} introduced a further generalization by incorporating an explicit coupling between the Ricci scalar $R$ and the trace of the energy-momentum tensor $T$. This coupling allows matter to directly affect the geometry beyond the standard minimal coupling assumed in GR. Such models naturally induce a non-conservation of the energy-momentum tensor and open new avenues for understanding the interaction between geometry and matter. Extensions of this framework, such as $f(R,L\_m)$ gravity \cite{Harko2010}, further enrich the model space and allow for even more diverse cosmological behaviors.
In recent years, numerous extensions and applications of modified gravity theories  frameworks have been extensively investigated in \cite{Shekh2025a,Shekh2025b,Shekh2025c,Shekh2025cc,Shekh2025d,Shekh2024e,Shekh2024f,Shekh2024g,Shekh2024h,Shekh2024i,Shekh2024j,Shekh2024k,Shekh2024l,Shekh2024m,Shekh2024n}.


In recent years, further generalizations have been proposed by introducing additional scalars representing different geometric properties of spacetime. In this regard, Bakry and Ibraheem \cite{Bakry2023} proposed a novel extension by introducing the non-metricity-like term $\Sigma$, which may encode effects of both normal and strong gravity regimes, as well as potential anti-gravity features. This leads to the development of the $f(R,\Sigma,T)$ gravity theory, where the action functional depends simultaneously on the Ricci scalar $R$, the scalar $\Sigma$, and the trace $T$. Such a structure allows the gravitational dynamics to be influenced by both geometric and material components in a highly non-trivial manner. The field equations of $f(R,\Sigma,T)$ gravity are considerably more complex than in simpler modified gravity theories, offering a richer set of cosmological solutions and evolution histories.

In the present work, we do not attempt to study the full complexity of the general 
	$f(R,\Sigma,T)$ theory. Instead, we restrict ourselves to the selective form $f(R,\Sigma,T) = R + \Sigma + 2 \pi \eta T ,$ where $\eta$ is a free constant parameter. which is not merely guided by mathematical simplicity, but physically interpreted as the leading-order linear extension of General Relativity within the broader class of $f(R,\Sigma,T)$ 	gravity model. In modified  theories of gravity, such linear models are often adopted as first-order approximations to isolate the dominant contributions which arises from additional geometric and matter coupling before introducing the higher-order nonlinear terms. The inclusion of a scalar $\Sigma$ incorporate an effects of nonmetricity, which is naturally emerges in the framework of symmetric teleparallel gravity and provide alternative geometric description of gravitation beyond curvature-based approaches. Moreover, the linear dependence on the trace of the energy--momentum tensor $T$ introduces a direct matter--geometry coupling, similar in spirit to $f(R,T)$ gravity, allowing for a possible energy exchange between the matter and geometric sectors.  Therefore, the adopted model represents a minimal yet physically meaningful extension of standard gravity which enables a controlled investigation of the impact of nonmetricity and matter coupling on cosmological dynamics. Although, a more general nonlinear forms of $f(R,\Sigma,T)$ lead to richer phenomenology, the present study focuses on this simplified framework to ensure analytical tractability and to clearly identify deviations from the standard $\Lambda$CDM cosmology at the background level. Also, this particular form is chosen because it preserves the essential new features of the theory, such as the inclusion of both geometric contributions through $\Sigma$ and matter–geometry couplings via $T$. Also, this functional form reduces the complexity of the theory while retaining the essential modifications introduced by the coupling of matter and geometry. By adopting the spatially flat Friedmann-Robertson-Walker (FRW) metric, we derive the modified Friedmann field equations and analyze their solutions under various observational constraints. In particular, we reconstruct the Hubble parameter $H(z)$ using the $O_m(z)$ diagnostic \cite{Sahni2008,Sahni2014}, which offers a model-independent approach to probe dark energy dynamics and distinguish between different cosmological scenarios.

One of the key advantages of using the $O_m(z)$ diagnostic is its minimal sensitivity to systematic errors, as it relies directly on measurements of the Hubble parameter rather than integrated quantities like the luminosity distance \cite{Shafieloo2012}. Furthermore, it is important to emphasize that the $O_m(z)$ diagnostic~\cite{Sahni2008} provides a model--independent test of dark energy, remaining constant for $\Lambda$CDM. Motivated by alternative parameterization, such as power--law~\cite{Myrzakulov2023} and exponential forms~\cite{Samaddar2025}, we employ a logarithmic parameterization for $O_m(z)$, expressed as $O_m(z) = \alpha \ln(1+z) + \beta$, where $\alpha$ and $\beta$ are free parameters constrained through observational data such as Hubble parameter measurements (OHD) and the Pantheon+ supernova sample \cite{Scolnic2018, Moresco2020}. This parameterization provides flexibility to capture deviations from the standard $\Lambda$CDM behavior and enables a detailed investigation of the dark energy equation of state evolution. In comparison, the statefinder diagnostic ($r,s$) provides a higher-order geometrical tool that distinguishes $\Lambda$CDM as a fixed point 
$(r,s) = (1,0)$, while the $(\omega, \omega')$-plane offers insights into the dynamical nature of dark energy (freezing versus thawing behavior). Similarly, growth-rate based diagnostics probe the perturbative sector. Our logarithmic parameterization of $O_m(z)$ is complementary to these approaches: it efficiently 	captures deviations from $\Lambda$CDM in a model–independent manner while retaining a direct observational connection through $H(z)$.

The precise nature of dark energy remains one of the deepest unsolved mysteries in cosmology. Various theoretical models have been proposed to characterize its origin and dynamics. Quintessence models introduce a canonical scalar field minimally coupled to gravity, with a time-varying equation of state parameter in the range $-1 < \omega < -1/3$ \cite{Ratra1988}. Phantom models assume $\omega < -1$, which can lead to exotic future singularities known as the Big Rip \cite{Caldwell2002, Sami2004}. Quintom models allow for the equation of state parameter to cross the phantom divide ($\omega = -1$) during cosmic evolution \cite{Elizalde2004, Nojiri2003}. Other models such as k-essence \cite{Armendariz2000}, tachyonic fields \cite{Padmanabhan2002}, chameleon fields \cite{Khoury2004}, and generalized Chaplygin gas models \cite{Bento2002} have also been extensively studied in the literature. Observational constraints from WMAP9 \cite{Hinshaw2023}, Planck 2015 \cite{Ade2015}, and Planck 2018 \cite{Planck2020} indicate that the present value of the dark energy equation of state lies very close to $\omega = -1$, but with some allowance for mild deviations consistent with quintessence or quintom behaviors. Our reconstructed equation of state parameter $\omega(z)$ remains within the quintessence regime throughout cosmic history, evolving smoothly toward $\omega = -1$ in the far future, thus offering a physically viable alternative to the cosmological constant. The present work is motivated by the need to explore more generalized modified gravity models that can naturally account for both the accelerated expansion of the Universe and the observational signatures of dynamical dark energy. The inclusion of both geometric ($\Sigma$) and material ($T$) corrections within $f(R,\Sigma,T)$ gravity offers a rich framework for investigating these questions. Furthermore, the flexibility of the logarithmic $O_m(z)$ parameterization allows us to fit the model to observational data without making rigid assumptions about the nature of dark energy.

This article is organized as follows: In Section II, we formulate the field equations for the $f(R,\Sigma,T)$ gravity model. In Section III, we present the reconstruction of the Hubble parameter using the $O_m(z)$ diagnostic and discuss the datasets employed. In Section IV, we analyze various cosmological parameters such as the equation of state parameter, energy conditions, stability criteria, and deceleration parameter. In Section V, we summarize our findings and conclusions. Throughout this work, we adopt natural units $8\pi G = c = 1$.
\section{Formulation}
\noindent In the framework of symmetric teleparallel gravity theory, the gravitation is described by the nonmetricity of space-time rather than curvature or torsion. The fundamental and the geometric quantities in this  formulation are metric tensor $g_{\mu\nu}$ and the independent affine connection $\Gamma^{\lambda}_{\ \mu\nu}$. The nonmetricity tensor is defined as
	\begin{equation}
		Q_{\lambda\mu\nu} = \nabla_{\lambda} g_{\mu\nu},
	\end{equation}
which measure the failure of metric to remain covariantly constant. From this, once construct the nonmetricity scalar $Q$ through suitable contraction of the nonmetricity tensor. In the present analysis, the scalar $\Sigma$ is introduced as a effective representation of the nonmetricity contributions, encapsulating the 	geometric effects arising from this sector. Hence, $\Sigma$ plays a role analogous to the Ricci scalar $R$ in curvature-based gravity, but within the nonmetricity framework. The teleparallel and extended symmetric teleparallel theories of gravity have been explored for several decades as alternatives to General Relativity (GR), offering a geometric framework in which gravitation is described using torsion or nonmetricity rather than curvature. Foundational works on teleparallel gravity date back to the 1960s–1980s, and comprehensive modern reviews are available~\cite{Aldrovandi2013, Krssak2018, Bahamonde2021}. Any contemporary formulation should be contextualized within this historical framework rather than attributing the first constructions to recent studies~\cite{Bakry2023}.
The action principle used to derive the field equations is written as
	\begin{small}
		\begin{equation}\label{e1}
			I = \frac{1}{16\pi} \int_{\Omega} \sqrt{-g} \left( B + \mathcal{L}_m \right) d^4x 
			= \frac{1}{16\pi} \int_{\Omega} \sqrt{-g} \left( g^{ab} B_{ab} + \mathcal{L}_m \right) d^4x,
		\end{equation}
	\end{small}
	where $\Omega$ denotes the space-time domain under consideration. The quantity $B_{ab}$ 
	represents a generalized geometric tensor constructed from both curvature and additional 
	geometric contributions, expressed schematically as $B_{ab} = R_{ab} + \Sigma_{ab}$. 
	Here, $R_{ab}$ is the usual Ricci tensor, while $\Sigma_{ab}$ encodes corrections arising 
	from extended geometric effects beyond standard Riemannian geometry.
	In the context of $f(R,\Sigma,T)$ gravity, the term $\Sigma_{ab}$ is modeled as 
	$\Sigma_{ab} = b\,\Phi_{ab}$, where $b$ is a dimensionless parameter and $\Phi_{ab}$ 
	represents an effective tensor capturing non-standard geometric contributions such as 
	nonmetricity or related extensions. The parameter $b$ plays a crucial role in controlling 
	the relative strength of these additional geometric effects with respect to the standard 
	curvature sector. 
		From a physical standpoint, $b$ can be interpreted as regulating the balance between 
	attractive gravitational effects and effective repulsive contributions that may drive 
	accelerated expansion. Different regimes of $b$ correspond to distinct cosmological 
	behaviors: values of $b$ less than unity enhance the influence of the additional geometric 
	sector, which may effectively act as a repulsive component, whereas larger values of $b$ 
	tend to favor the dominance of curvature-driven attraction. The case where the additional 
	contributions become negligible corresponds to a recovery of standard General Relativity 
	as an effective limit.
		With these considerations, the generalized action for $f(R,\Sigma,T)$ gravity can be 
	written as
	\begin{equation}\label{e2}
		I = \frac{1}{16\pi} \int_{\Omega} \sqrt{-g} \left( f(R, \Sigma, T) + \mathcal{L}_m \right) d^4x.
\end{equation}
where $R$ is the Ricci scalar, $\Sigma$ represents contributions from the nonmetricity scalar, $T$ is the trace of the matter energy–momentum tensor, and $\mathcal{L}_m$ is the matter Lagrangian. The introduction of additional degrees of freedom in $f(R,\Sigma,T)$ gravity is motivated by the need to capture effects that are inaccessible to pure curvature-based modifications such as $f(R)$ gravity. The scalar $\Sigma$ originates from the non-metricity sector of symmetric teleparallel geometry, which enriches the geometric structure of the theory and permits the inclusion of both weak- and strong-field corrections. On the other hand, the explicit dependence on the trace of the energy–momentum tensor $T$ introduces a non-minimal matter–geometry coupling, thereby allowing direct exchange of energy between spacetime and matter sectors. Such a feature, also present in $f(R,T)$ gravity, has been shown to provide new phenomenological possibilities for explaining cosmic acceleration and entropy production. Together, the $\Sigma$ and $T$ contributions extend the predictive capacity of the model beyond that of standard $f(R)$ or $f(T)$ theories.\\

\noindent The variation of this action with respect to the metric yields the field equations
\begin{small}
\begin{equation}\label{eq2}
\begin{split}
R_{ab} \frac{\partial f}{\partial R} &+ \Sigma_{ab} \frac{\partial f}{\partial \Sigma} - \frac{1}{2} g_{ab} f \\
&+ \left(g_{ab} \nabla^\gamma \nabla_\gamma - \nabla_a \nabla_b \right) \frac{\partial f}{\partial R} = 8\pi T_{ab} + \left(T_{ab} + p g_{ab}\right) \frac{\partial f}{\partial T}.
\end{split}
\end{equation}
\end{small}

\noindent However, this only captures part of the system: the connection degrees of freedom associated with the nonmetricity scalar also generate independent field equations. A full treatment requires introducing the scalar invariants explicitly and varying the action with respect to both the metric and the connection. Neglecting these contributions means that the analysis does not account for all degrees of freedom of the underlying geometry. Furthermore, the potential non-conservation of energy–momentum in $f(R, \Sigma, T)$ gravity is not addressed, leaving open questions regarding energy exchange between geometry and matter.

The energy–momentum tensor for matter is defined as
\begin{equation}\label{4}
T_{ab} = (\rho + p) u_a u_b + p g_{ab},
\end{equation}
where $\rho$ is the energy density of matter, $p$ is
the pressure of the matter, and $u_a$ is the fluid-four velocity vector, where $u_a u^b =-1$. In this article, we assume that the function $f(R, \Sigma, T)$ is
given by
\begin{equation}\label{5}
f(R, \Sigma, T) = R + \Sigma + 2\pi \eta T.
\end{equation}
where $\eta$ is a constant parameter, the resulting field equations reduce to a rescaled version of the standard Friedmann equations under a perfect fluid equation of state. The choice of the functional form (\ref{5})
	is motivated as a minimal linear extension of General Relativity within the broader 
	$f(R,\Sigma,T)$ framework. This form retains the essential physical ingredients of the 
	theory, namely curvature ($R$), nonmetricity effects ($\Sigma$), and matter--geometry 
	coupling ($T$), while avoiding the additional complexities associated with nonlinear 
	models. Such linear approximations are commonly employed in modified gravity to isolate 
	leading-order deviations from standard cosmology before considering higher-order 
	corrections.
	We note that more general nonlinear forms of $f(R,\Sigma,T)$ could indeed probe richer 
	coupling effects; however, the present study focuses on the linear regime to ensure 
	analytical tractability and to clearly identify the role of each contribution in the 
	cosmic dynamics. While this form is analytically tractable, it limits the novelty of the cosmological dynamics, producing only rescaled matter contributions relative to GR. Importantly, this functional choice produces a rescaled version of the standard Friedmann equations that mimic the $\Lambda$CDM background expansion, thereby providing a well–behaved and observationally consistent cosmological 
dynamics. This balance between theoretical richness and mathematical simplicity is the main rationale for adopting the present model. Also, it is important to emphasize that in this generalized framework, both the metric and the affine connection contribute to the gravitational dynamics. While the metric determines distances and causal structure, the connection encodes the geometric properties such as nonmetricity. In a complete formulation, the action should be varied with respect to both the metric and the connection. In the present work, we adopt a simplified approach in which the effective scalar $\Sigma$ captures the dominant contributions of the nonmetricity sector, allowing us to focus on the cosmological implications of the model.\\ 
\noindent Using the function (\ref{5}), the final form of $f(R, \Sigma, T)$ equation (\ref{eq2}) is given by
\begin{equation}\label{6}
R_{ab} + \Sigma_{ab} - \frac{1}{2} g_{ab} (R + \Sigma) = 2(4\pi + \pi \eta) T_{ab} + \pi \eta g_{ab}(T + 2p)
\end{equation}
which can be alternatively expressed as
\begin{small}
\begin{equation}\label{7}
	B_{ab} = R_{ab} + \Sigma_{ab} = 2(4\pi + \pi \eta) \left(T_{ab} - \frac{1}{2} g_{ab} T\right) - \pi \eta g_{ab} (T + 2p),
\end{equation}
\end{small}
It is instructive to recover the standard $\Lambda$CDM model as a limiting case of the present framework. In the absence of nonmetricity and matter--geometry coupling, i.e., $\Sigma \rightarrow 0$ and $\eta \rightarrow 0$, the function reduces 
to $f(R,\Sigma,T) \rightarrow R$, and the standard Einstein field equations are 
recovered. Consequently, the Friedmann equations reduce to their usual form, $	H^2 = \frac{8\pi G}{3}\rho + \frac{\Lambda}{3}$ which describes the $\Lambda$CDM cosmology. This demonstrates that the present model 
is a consistent extension of standard cosmology.
To facilitate the solution of field equations in $f(R, \Sigma, T)$ extended symmetric teleparallel gravity, simplifying assumptions are often necessary. In this work, we adopt the homogeneous, isotropic, and spatially flat Friedmann-Robertson-Walker (FRW) metric, given by:

\begin{equation}\label{8}
ds^{2}=-dt^{2}+\delta_{ij} g_{ij} dx^{i} dx^{j},{\;\;\;\;} i,j=1,2,3,.....N,
\end{equation}
This choice of metric allows us to explore the cosmological implications of $f(R, \Sigma, T)$ gravity in a straightforward and analytically tractable manner.
where $g_{ij}$ are the function of $(-t, x^{1}, x^{2}, x^{3})$ and $t$ refers to the cosmological/cosmic time measure in Gyr. In the four-dimensional FRW space-time, the equation above yields the following:
\begin{equation}\label{9}
	\delta_{ij} g_{ij}=a^{2}(t)
\end{equation} 
where $a$ be the average scale factor of the Universe and $t$ is the cosmic time in Gyr. The aforementioned relationships demonstrate that all three metrics are equivalent in the FRW universe (i.e $g_{11} = g_{22} = g_{33} =a^{2}(t)$). \\
By utilizing Equation (\ref{8}) for the space–time in a comoving coordinate system, we can obtain the components of $f(R, \Sigma, T)$ gravity field Equations from the equation (\ref{7}) as follows:
\begin{equation}\label{10}
	3(b-1)\left(\dot{H}+H^2\right)=4 \pi \rho+4 \pi \left(3+\eta\right)p
\end{equation}
\begin{equation}\label{11}
	3(1-b)^2 H^2= \pi \left(8+3\eta\right) \rho - \pi \eta p
\end{equation}
 Here, the parameter $b$ is a dimensionless constant that arises from the assumed relation between the geometric scalars or from the parametrization adopted to simplify the field equations. It effectively controls the deviation from the standard Friedmann dynamics and plays a role in determining the evolution of the Hubble parameter and related cosmological quantities.
Moreover, it is worthwhile to note that Eqs. (\ref{10})–(\ref{11}) do not, in general, imply the usual energy-conservation equation
\[
\dot\rho+3H(\rho+p)=0,
\]
except for special choices of the parameters or special evolutions \(H(t)\). To explained this, we consider the following equation
\begin{equation}\label{eq-Q}
\mathcal{Q}\equiv\dot\rho+3H(\rho+p),
\end{equation}
Eqs. (\ref{10}), (\ref{11}) and (\ref{eq-Q}) lead
\begin{widetext}
\begin{equation}\label{eq-Q1}
\mathcal{Q}
=\frac{(b-1)}{4\pi(\eta+2)(\eta+4)}
\Big[\,12b(\eta+2)H^3+(8b\eta+24b+6\eta)H\dot H+\eta\,\ddot H\Big].
\end{equation}
\end{widetext}
Hence the energy conservation law \(\mathcal{Q}=0\) holds only if the right-hand side of Eq. (\ref{eq-Q1}) vanishes. Mathematically, this is possible for two cases: i) $b = 1$ \& ii) $12b(\eta+2)H^3+(8b\eta+24b+6\eta)H\dot H+\eta\,\ddot H = 0$. But \(b=1\) also makes the original field equations degenerate therefore, one must omit \(b=1\). Thus, the conservation can be restored for special cosmic evolutions $H(t)$ obeying the differential constraint. It is important to stress that in the general $f(R,\Sigma,T)$ framework the 
energy–momentum tensor is not, in general, covariantly conserved, i.e. 
$\nabla_\mu T^{\mu\nu} \neq 0$ \cite{Rastall1972,Harko2011}. In the present selective model this feature 
appears explicitly through Eq.~(12), where the conservation law can only be 
restored for special choices of $H(t)$. Rather than being a drawback, such 
non-conservation can be interpreted as an exchange of energy between matter 
and the underlying geometric sector. In cosmology, this has several possible 
interpretations: it may correspond to effective particle creation processes 
in an expanding universe, to entropy production in non-equilibrium 
thermodynamics, or to an effective bulk viscous contribution. Similar 
mechanisms appear in $f(R,T)$ gravity and in Rastall gravity, where the usual 
conservation law is generalized. Thus, the lack of strict conservation is a 
structural property of the theory and can be exploited to describe dissipative 
or non-equilibrium effects in cosmic evolution.\\

\noindent The expressions of $p$ and $\rho$  are obtained by solving Eqs. (\ref{10}) and (\ref{11}):
\begin{equation}\label{12}
p = \frac{1-b}{4(8\pi + 6\eta +\pi \eta^2)} \left[ (8+3\eta) \dot{H} + (4(3-b) + 3\eta)H^2 \right]
\end{equation}
\begin{equation}\label{13}
\rho = \frac{1-b}{4(8\pi^2 + 6\pi \eta + \eta^2)} \left[ -\eta \dot{H} + (12\pi(1-b) + \eta(3 - 4b))H^2 \right]
\end{equation}
\section{Method of parameterization}
Obtaining explicit solutions to the field equations and subsequently analyzing the universe's physical behavior necessitates making supplementary assumptions. While various analytical methods exist for understanding cosmic evolution, one approach involves parameterizing specific parameters. The \( O_m(z) \) diagnostic serves as a valuable tool in cosmology for examining the nature of dark energy (DE) and differentiating between a constant dark energy density (cosmological constant, \(\Lambda\)) and models where DE density evolves dynamically. Introduced as a model-independent test for \(\Lambda\) \cite{Sahni2008, Sahni2014}, it offers an alternative to methods that rely on specific DE models. Unlike the equation of state parameter \( \omega_{DE} \), which is sensitive to assumptions about the evolution of DE density, \( O_m(z) \) is constructed from the Hubble parameter \( H(z) \) and the redshift \( z \). Since \( H(z) \) can be determined through observations such as supernovae of Type Ia (SNe Ia), baryon acoustic oscillations (BAO), and Hubble parameter measurements, \( O_m(z) \) provides a robust, observationally linked way to probe DE's characteristics. The mathematical definition of the \( O_m(z) \) diagnostic is given by:
\begin{equation}\label{14}
O_m(z) = \frac{H(z)^2 - H_0^2}{H_0^2(1+z)^3 - H_0^2}
\end{equation}
where \( H_0 \) is the Hubble constant at present (\( z = 0 \)). \\

\noindent A key feature of the \( O_m(z) \) diagnostic is its ability to distinguish between different DE behaviors. A constant \( O_m(z) \) value across redshifts, equal to the present-day matter density parameter \( \Omega_{0m} \), indicates a cosmological constant (\( \omega = -1 \)). In contrast, a redshift-dependent \( O_m(z) \) suggests dynamical DE. Furthermore, the sign of the \( O_m(z) \)  provides information about the type of dynamical DE: The positive sign suggests a phantom DE scenario where \( \omega < -1 \) whereas the negative sign  indicates a quintessence-like DE regime with \( \omega > -1 \). To encompass a wider spectrum of potential DE behaviors, we propose a parameterization of \( O_m(z) \) using a logarithmic function, 

The standard $O_m(z)$ diagnostic, given in Eq.~(\ref{14}), remains constant and equal to 
	$\Omega_{0m}$ for the $\Lambda$CDM scenario. In order to capture potential 
	departures from this constant behavior, we introduce the generalized logarithmic 
	form
\begin{equation}\label{15}
Om_(z) = \alpha \ln(1+z) + \beta ,
\end{equation}
where $\alpha$ and $\beta$ are free parameters. This functional extension is motivated by the fact that logarithmic terms naturally arise in many cosmological parameterizations of the deceleration parameter $q(z)$ and the equation of state $\omega(z)$, and provide a smooth interpolation between early- 
and late-time regimes. The two-parameter form allows one to probe dynamical 
dark energy in a model–independent manner while still maintaining analytical 
simplicity. A key property of this parameterization is that the special case $\alpha = 0$ 
reduces to $Om(z) = \beta$, a constant. Identifying $\beta$ with the present-day 
matter density parameter $\Omega_{0m}$, one recovers the $\Lambda$CDM diagnostic 
as in Eq.~(\ref{14}). Therefore, the proposed form retains the correct $\Lambda$CDM 
limit and extends it only when $\alpha \neq 0$. This makes the new 
parameterization a controlled generalization of the standard $O_m(z)$ diagnostic.

In our quest to understand the nature of dark energy (DE) and potential departures from the standard cosmological constant (\(\Lambda\)), we introduce a logarithmic parameterization for the \(O_m(z)\) diagnostic. This parameterization is defined by the free parameters \(\alpha\) and \(\beta\), which will be determined through comparisons with observational data. The adoption of a logarithmic evolution for \(O_m(z)\) provides a versatile yet physically grounded approach to investigate DE. Notably, a logarithmic dependence on redshift \(z\) arises naturally in various physical scenarios, including parameterizations of the deceleration parameter and the equation of state (EoS) parameter. 
This functional form is well-suited for capturing gradual and smooth variations in cosmic evolution across different redshifts, making it appropriate for describing subtle deviations from the \(\Lambda\)CDM model.

Physically, the parameter \(\beta\) establishes the fundamental value of \(O_m(z)\). In the specific case where \(\alpha\) equals zero, our model simplifies to a constant \(O_m(z) = \beta\), a scenario that aligns with a cosmological constant (\(\omega = -1\)), implying a dark energy density that remains constant over cosmic time. When \(\alpha\) is non-zero, it dictates the rate at which \(O_m(z)\) deviates from this constant value, offering a direct way to probe the evolution of dark energy. A positive value of \(\alpha\) corresponds to a phantom-like behavior of dark energy (\(\omega < -1\)), suggesting a scenario where the energy density of dark energy increases with time, potentially leading to a 'Big Rip' scenario in the distant future. Conversely, a negative value of \(\alpha\) indicates quintessence-like dynamics (\(\omega > -1\)), where the dark energy density decreases with time, similar to a scalar field rolling down a potential. Given that observations of the Hubble parameter \(H(z)\) generally suggest a mild and gradual evolution of dark energy with redshift, the logarithmic form provides a phenomenological description that can effectively capture such trends without imposing overly restrictive assumptions on the underlying physics. Overall, this two-parameter logarithmic form naturally accommodates deviations from a constant \(O_m(z)\) (as predicted by \(\Lambda\)CDM) while ensuring a smooth transition between the early and late stages of the universe's evolution. The parameter \(\alpha\) thus serves as a key indicator of the evolving nature of dark energy, while \(\beta\) provides a baseline related to the present-day matter content (\(\Omega_{0m}\)) within the standard cosmological framework.\\

\noindent Eqs. (\ref{14}) and (\ref{15}) lead to
\begin{equation}
\label{H}
H(z) = H_{0} \Big[ (1+z)^3 + \big((1+z)^3 - 1\big) \, (\alpha \ln(1+z) + \beta) \Big]^{\frac{1}{2}}
\end{equation}
It is important to emphasize that Eq.~(\ref{H}), obtained from the logarithmic 
	$O_m(z)$, represents a phenomenological parameterization of the Hubble parameter. 
	It is not a closed-form solution of the $f(R,\Sigma,T)$ field equations, but 
	rather an observationally motivated ansatz designed to test the model against 
	data. Similar phenomenological approaches are widely employed in cosmology, 
	for example the Chevallier–Polarski–Linder (CPL) parameterization of the dark 
	energy equation of state. The advantage is that the resulting $H(z)$ form is 
	directly constrained by data, while the theoretical connection to 
	$f(R,\Sigma,T)$ gravity is encoded through the interpretation of the 
	parameters and their consistency with the background equations derived in 
	Section II.

\noindent For flat $\Lambda$CDM model, $H(z)$ is read as
\begin{equation}
\label{H-L}
H(z) = H_{0}\Big[\Omega_{m}(1+z)^{3} + \Omega_{\Lambda} \Big]^{\frac{1}{2}}
\end{equation}
where $\Omega_{m}$ and $\Omega_{\Lambda}$ stand for matter density parameter and dark energy density parameter respectively.
We note that the logarithmic $O_m(z)$ formulation does not explicitly feature the matter density parameter $\Omega_{0m}$, and thus cannot by itself be employed 	to describe perturbations or the full growth of cosmic structures. The present analysis is restricted to background cosmology, and the omission of $\Omega_{0m}$ should be regarded as a limitation of the diagnostic. A complete perturbative treatment of the $f(R,\Sigma,T)$ framework would require incorporating the density parameter explicitly, which we leave for future work. Nonetheless, the background analysis remains valuable for distinguishing dynamical dark energy from $\Lambda$CDM and for providing constraints from current $H(z)$ and SNIa datasets.
\begin{center}
\bf{Data sets}
\end{center}

A comprehensive analysis of the dynamics and the overall validity of this theoretical framework necessitates the determination of its free parameters specifically \(\alpha\), \(\beta\), and \(H_0\)  through rigorous comparison with observational data. Datasets such as observational Hubble data, and the Pantheon+ compilation will be crucial for this purpose, as they offer accurate measurements of the Hubble parameter \(H(z)\) and other relevant cosmological observables.\\
\subsection{The Observational $H(z)$ data (OHD)}
For constraining the model parameters, we have employed the 77 uncorrelated Hubble observations $H(z)$ that lie in the redshift range $0 \leq z \leq 2.36$. These $77$ $H(z)$ data points with its original references are given in following table I. To assure the probing of parameters and account for systematic effects in the Cosmic Chronometer (CC) data, we admit entire covariance matrix as provided by~\cite{Moresco2020}. More details of CC covariance gauge are available at \url{https://gitlab.com/mmoresco/CCcovariance}. The chi-squared function to comprise the covariance matrix as follows:
\[
\chi^2_{\mathrm{OHD}} = (\mathbf{H}_{\text{obs}} - \mathbf{H}_{\text{model}})^T \cdot \mathbf{C}^{-1} \cdot (\mathbf{H}_{\text{obs}} - \mathbf{H}_{\text{model}}),
\]
where $\mathbf{H}_{\text{obs}}$ and $\mathbf{H}_{\text{model}}$ denote the observed and theoretical $H(z)$ values, respectively, and $\mathbf{C}$ represents the covariance matrix. Moreover, the consolidation of CC data with a covariance matrix enhances the statistical rigor of our analysis ~\cite{Moresco2020}.\\
\begin{table*}
\caption{The 77 Hubble Parameter Data from $H(z)$ measurements used in the present analysis in units of $\mathrm{km\,s^{-1}Mpc^{-1}}$. Method (I) corresponds to the Cosmic Chronometric method, method (II) to $BAO$ signal in galaxy distribution, and method (III) to $BAO$ signal in Ly$\alpha$ forest distribution alone, or cross-correlated with $QSOs$.}
\centering
\begin{tabular}{ccccc|ccccc}
\hline\hline
S.No. & $z$ & $H(z)$ & Method & Ref. & S.No. & $z$ & $H(z)$ & Method & Ref. \\
\hline
1 & 0.00 & 69.1 $\pm$ 1.3 & I & \cite{Farooq} & 40 & 0.68 & 93.9 $\pm$ 8.1 & I & \cite{Moresco} \\
2 & 0.07 & 70.4 $\pm$ 20 & I & \cite{Zhang} & 41 & 0.73 & 99.3 $\pm$ 7.1 & I & \cite{Blake} \\
3 & 0.07 & 69.0 $\pm$ 19.6 & I & \cite{Zhang} & 42 & 0.7812 & 105.0 $\pm$ 12 & I & \cite{Moresco} \\
4 & 0.09 & 70.4 $\pm$ 12.2 & I & \cite{Simon} & 43 & 0.781 & 107.1 $\pm$ 12.2 & I & \cite{Moresco} \\
5 & 0.10 & 70.4 $\pm$ 12.2 & I & \cite{Zhang} & 44 & 0.875 & 127.6 $\pm$ 17.3 & I & \cite{Moresco} \\
6 & 0.12 & 68.6 $\pm$ 26.2 & I & \cite{Zhang} & 45 & 0.8754 & 125.0 $\pm$ 17 & I & \cite{Moresco} \\
7 & 0.12 & 70.0 $\pm$ 26.7 & I & \cite{Farooq} & 46 & 0.88 & 91.8 $\pm$ 40.8 & I & \cite{Stern} \\
8 & 0.17 & 83.0 $\pm$ 8 & I & \cite{Simon} & 47 & 0.880 & 90.0 $\pm$ 40 & I & \cite{Ratsimbazafy} \\
9 & 0.17 & 84.7 $\pm$ 8.2 & I & \cite{Simon} & 48 & 0.90 & 69.0 $\pm$ 12 & I & \cite{Simon} \\
10 & 0.179 & 76.5 $\pm$ 4 & I & \cite{Moresco} & 49 & 0.90 & 119.4 $\pm$ 23.4 & I & \cite{Simon} \\
11 & 0.1791 & 75.0 $\pm$ 4 & I & \cite{Moresco} & 50 & 0.900 & 117.0 $\pm$ 23 & I & \cite{Simon} \\
12 & 0.199 & 76.5 $\pm$ 5.1 & I & \cite{Moresco} & 51 & 1.037 & 157.2 $\pm$ 20.4 & I & \cite{Moresco} \\
13 & 0.1993 & 75.0 $\pm$ 5 & I & \cite{Moresco} & 52 & 1.037 & 154.0 $\pm$ 20 & I & \cite{Moresco} \\
14 & 0.20 & 72.9 $\pm$ 29.6 & I & \cite{Zhang} & 53 & 1.30 & 171.4 $\pm$ 17.3 & I & \cite{Simon} \\
15 & 0.20 & 74.4 $\pm$ 30.2 & I & \cite{Zhang} & 54 & 1.300 & 168.0 $\pm$ 17 & I & \cite{Simon} \\
16 & 0.24 & 81.5 $\pm$ 2.7 & II & \cite{Gaztanaga} & 55 & 1.363 & 160.0 $\pm$ 33.6 & a & \cite{Moresco2} \\
17 & 0.27 & 78.6 $\pm$ 14.3 & I & \cite{Simon} & 56 & 1.363 & 163.3 $\pm$ 34.3 & I & \cite{Moresco2} \\
18 & 0.28 & 88.8 $\pm$ 36.3 & I & \cite{Zhang} & 57 & 1.43 & 177.0 $\pm$ 18 & I & \cite{Simon} \\
19 & 0.28 & 90.6 $\pm$ 37.3 & I & \cite{Farooq} & 58 & 1.43 & 180.6 $\pm$ 18.3 & I & \cite{Simon} \\
20 & 0.35 & 84.4 $\pm$ 8.6 & II & \cite{Farooq} & 59 & 1.53 & 140.0 $\pm$ 14 & I & \cite{Simon} \\
21 & 0.3519 & 83.0 $\pm$ 14 & I & \cite{Moresco} & 60 & 1.53 & 142.9 $\pm$ 14.2 & I & \cite{Simon} \\
22 & 0.352 & 84.7 $\pm$ 14.3 & I & \cite{Moresco} & 61 & 1.75 & 202.0 $\pm$ 40 & I & \cite{Simon} \\
23 & 0.38 & 81.5 $\pm$ 1.9 & II & \cite{Alam} & 62 & 1.75 & 206.1 $\pm$ 40.8 & I & \cite{Simon} \\
24 & 0.3802 & 83.0 $\pm$ 13.5 & I & \cite{Moresco1} & 63 & 1.965 & 186.5 $\pm$ 50.4 & I & \cite{Moresco2} \\
25 & 0.3802 & 84.7 $\pm$ 14.1 & I & \cite{Moresco1} & 64 & 1.965 & 190.3 $\pm$ 51.4 & I & \cite{Moresco2} \\
26 & 0.40 & 95.0 $\pm$ 17 & I & \cite{Simon} & 65 & 2.30 & 228.0 $\pm$ 8.1 & III & \cite{Delubac} \\
27 & 0.40 & 96.9 $\pm$ 17.3 & I & \cite{Simon} & 66 & 2.34 & 226.5 $\pm$ 7.1 & III & \cite{Delubac} \\
28 & 0.4004 & 77.0 $\pm$ 10.2 & I & \cite{Moresco1} & 67 & 2.36 & 230.6 $\pm$ 8.2 & III & \cite{Font-Ribera} \\
29 & 0.4004 & 78.6 $\pm$ 10.4 & I & \cite{Moresco1} & 68 & 0.4247 & 87.1 $\pm$ 11.2 & I & \cite{Moresco1} \\
30 & 0.4247 & 87.1 $\pm$ 11.2 & I & \cite{Moresco1} & 69 & 0.4247 & 88.9 $\pm$ 11.4 & I & \cite{Moresco1} \\
31 & 0.4247 & 88.9 $\pm$ 11.4 & I & \cite{Moresco1} & 70 & 0.4497 & 92.8 $\pm$ 12.9 & I & \cite{Moresco1} \\
32 & 0.43 & 88.3 $\pm$ 3.8 & I & \cite{Farooq} & 71 & 0.4497 & 94.7 $\pm$ 13.1 & I & \cite{Moresco1} \\
33 & 0.44 & 84.3 $\pm$ 7.9 & I & \cite{Blake} & 72 & 0.470 & 89.0 $\pm$ 34.0 & I & \cite{Ratsimbazafy} \\
34 & 0.4497 & 92.8 $\pm$ 12.9 & I & \cite{Moresco1} & 73 & 0.47 & 90.8 $\pm$ 50.6 & I & \cite{Ratsimbazafy} \\
35 & 0.4497 & 94.7 $\pm$ 13.1 & I & \cite{Moresco1} & 74 & 0.4783 & 80.0 $\pm$ 99.0 & I & \cite{Moresco1} \\
36 & 0.470 & 89.0 $\pm$ 34.0 & I & \cite{Ratsimbazafy} & 75 & 0.4783 & 82.5 $\pm$ 9.2 & I & \cite{Moresco1} \\
37 & 0.47 & 90.8 $\pm$ 50.6 & I & \cite{Ratsimbazafy} & 76 & 0.48 & 99.0 $\pm$ 63.2 & I & \cite{Ratsimbazafy} \\
38 & 0.4783 & 80.0 $\pm$ 99.0 & I & \cite{Moresco1} & 77 & 0.64 & 98.82 $\pm$ 2.98 & II & \cite{Wang1} \\
39 & 0.4783 & 82.5 $\pm$ 9.2 & a & \cite{Moresco1} &  &  &  &  &  \\
\hline\hline
\end{tabular}
\end{table*}

\subsection{Pantheon Plus (Pantheon+)}
The distance modulus measurements of type Ia supernovae (SN Ia) were obtained using the Pantheon plus sample \cite{Scolnic2018}. The redshift range $z \in [0.001, 2.26]$ has 1701 light curves representing 1550 SN Ia data points. The investigation of the expansion rate heavily relies on SNe Ia.\\ 
The luminosity distance is read as
\begin{equation}
D_{L}(z) = (1+z)\int^{z}_{0}\frac{H_{0}}{H(z^{\prime})}dz^{\prime} ,
\end{equation}
Thus, the distance modulus is computed as
\begin{equation}
\mu(z) = 5 \log_{10} \left(\frac{D_L(z)}{~1{Mpc}}\right) + 25. 
\end{equation}
Furthermore, it is worthwhile to note that the Pantheon+ sample provides distance modulus $\mu(z)$, but these are not absolutely calibrated because the absolute magnitude $M_{B}$ of Type Ia supernovae is not fixed within the dataset itself and so $M_B$ needs to be added to the distance modulus to calculate the theoretical values for $m_B$ for each SN
\begin{equation}
m^{\rm theory}_{B,i} = \mu(z_i) + M_B.
\end{equation}
Thus, the $\chi^2$ takes the following form
\begin{equation}
\chi^2 = \sum_{i,j} (m^{\rm data}_{B,i} - m^{\rm theory}_{B,i})^{T}  \mathcal{C}_{i,j}^{-1} (m^{\rm data}_{B,j} - m^{\rm theory}_{B,j})
\end{equation}
where $i, j$ span the set of SN in the dataset and $\mathcal{C}$ is the Pantheon+ covariance matrix.\\
\begin{figure}
\centering
\includegraphics[scale=0.60]{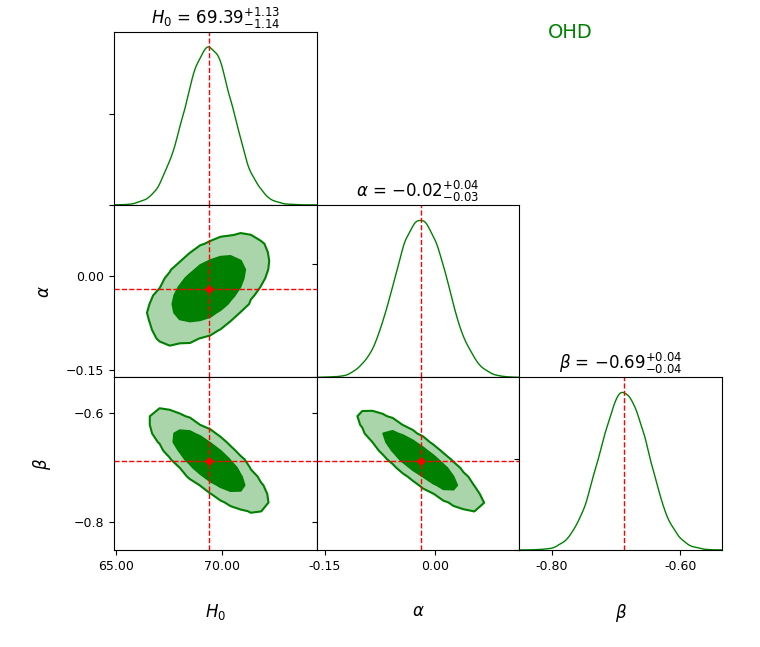}
\caption{One-dimensional marginalized distribution and two-dimensional contours at $1\sigma$ and $2\sigma $ confidence levels of proposed model by using the most recent 77 observational Hubble data to constrain model parameters. The unit of $H_{0}$ is $\;km\;s^{-1}\;Mpc^{-1}$}.\label{OHDF1}
\end{figure}  
\begin{figure}
\centering
\includegraphics[scale=0.60]{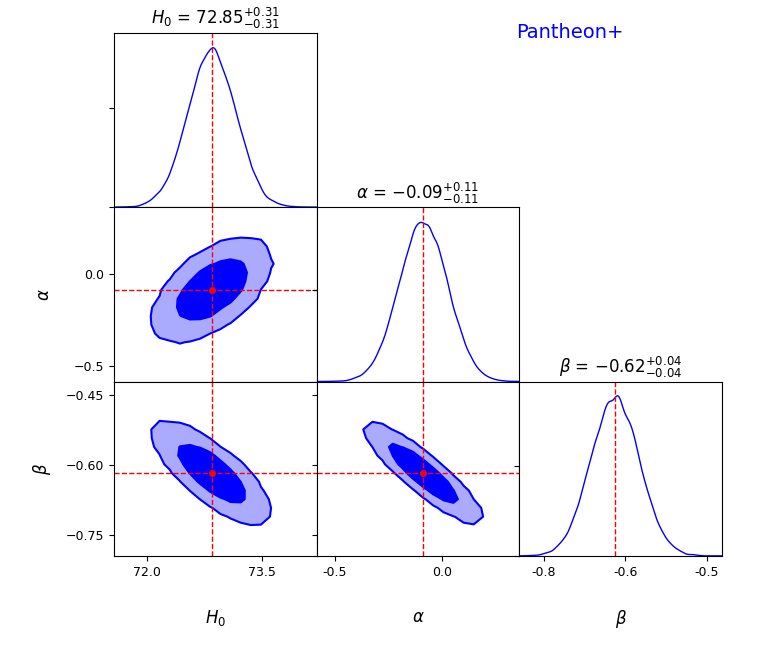}
\caption{One-dimensional marginalized distribution and two-dimensional contours at $1\sigma$ and $2\sigma $ confidence levels of proposed model by using the Pantheon plus data sets to constrain the model parameters. The unit of $H_{0}$ is $\;km\;s^{-1}\;Mpc^{-1}$}.\label{PPF2}
\end{figure}  

\noindent For the purpose of evaluating the model parameters $H_{0}$, $\alpha$ \& $\beta$ of this model, observational data and statistical approaches may be used. Let $E_{obs}$ represent the values that have been observed and let $E_{th}$ represent the values that have been theoretically calculated to conform to the model of the universe. Estimating the parameters of the model can be achieved by using statistical techniques to compare the two values $E_{obs}$ and $E_{th}$ within the model. It is possible to interpret the formulation of the $\chi^2$ estimator as
\begin{equation}
\label{chi1}
\chi^{2} = \sum_{i=1}^{N}\left[\frac{E_{th}(\Phi_{c},z_{i})- E_{obs}(\Phi_{l},z_{i})}{\sigma_{i}}\right]^{2}
\end{equation}
In this context, the symbols $E_{th}(\Phi_{c},z_{i})$ and $E_{obs}(\Phi_{l},z_{i})$ represent the theoretical values and the observed values of the relevant parameters by themselves. The $\Phi_{c}$ is the cosmological parameter sets, and $\Phi_{c} = (H_{0}, \alpha, \beta)$. The $\Phi_{l}$ denotes the nuisance parameter sets and we have $\phi_{l} = (H_{0}, \alpha, \beta)$ for OHD and Pantheon plus sample respectively. Further, we refer to the standard errors in $E_{obs}(z_{i})$ as $\sigma_{i}$ and the number of data points as N.\\

\noindent Fig. \ref{OHDF1} depicts the two-dimensional confidence contours at $1\sigma$ and $2\sigma$ intervals of our model using the recent 77 OHD dataset while Fig. \ref{PPF2} represents a view of the $1\sigma$ and $2\sigma$ confidence regions of our model using Pantheon plus data sets. The numerical results are in table II.
\begin{widetext}
\begin{table}[h!]
\centering
\setlength{\tabcolsep}{10pt}
\caption{Comparison of best-fit parameters between our model and the standard flat $\Lambda$CDM model using OHD and Pantheon+ datasets.}
\label{tab:model_comparison}
\begin{tabular}{lcccccc}
\toprule
Datasets~~~~~ &~~~~~ Model~~~ &~~~ $H_{0}$ (km s$^{-1}$ Mpc$^{-1}$)~~~ &~~~ $\alpha$ ~~~&~~~ $\beta$ ~~~~ & ~~~~ $\Omega_{m}$ ~~~~ & ~~~~ $\Omega_{\Lambda}$ \\[6pt]
\midrule
OHD & Our model     & $69.39^{+1.13}_{-1.14}$ & $-0.02^{+0.04}_{-0.03}$ & $-0.69^{+0.04}_{-0.04}$  & $ - $ &  $ - $\\[6pt]
    & $\Lambda$CDM  & $69.72^{+0.97}_{-0.97}$ & $-$              & $-$ & $0.29^{+0.02}_{-0.02}$ & $0.71^{+0.02}_{-0.02}$  \\ [6pt]
\addlinespace[6pt]
Pantheon+ & Our model     & $72.85^{+0.31}_{-0.31}$ & $-0.09^{+0.11}_{-0.11}$ & $-0.62^{+0.04}_{-0.04}$  & $ - $ & $ - $ \\[6pt]
          & $\Lambda$CDM  & $ 72.97^{+0.26}_{-0.26}$ & $-$              & $-$  & $0.35^{+0.03}_{-0.03}$   & $0.65^{+0.03}_{-0.03}$\\ [6pt]
\bottomrule
\end{tabular}
\end{table}
\end{widetext}
\begin{figure}[ht]
\centering
\includegraphics[scale = 0.50]{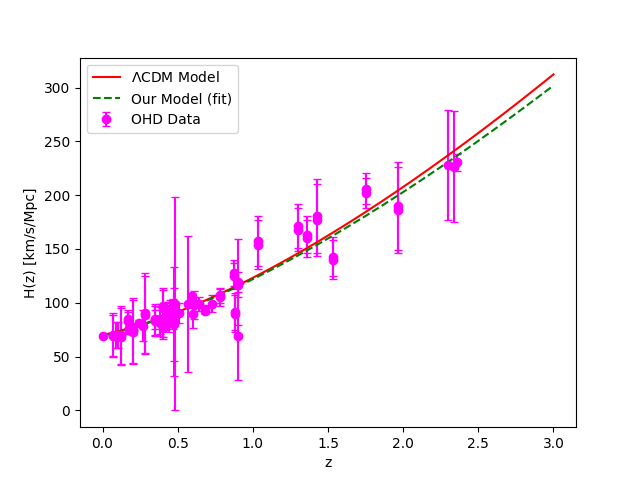}
\includegraphics[scale = 0.50]{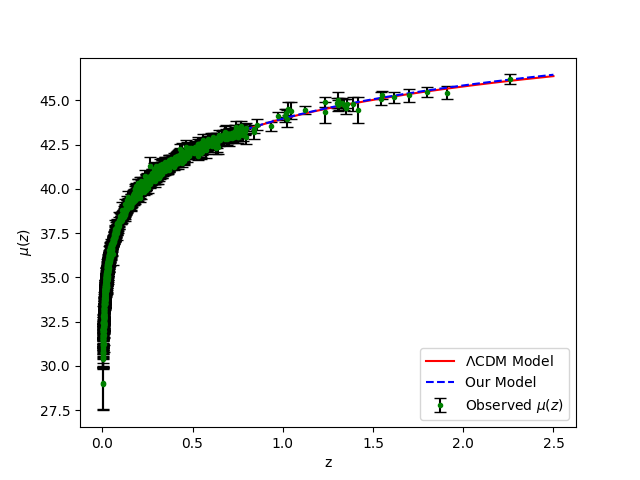}
\caption{The left panel of above figure shows the variation of H(z) of our model with redshift z and its comparison with $\Lambda$CDM model while the right panel of above figure exhibits the variation of distance modulus $\mu(z)$ of our model with redshift z and its comparison with $\Lambda$CDM model.}\label{FR1}
\end{figure}
The left panel of Fig.~\ref{FR1} shows the variation of $H(z)$ with redshift $z$ for our model, along with its comparison to the $\Lambda$CDM model, including the associated $1\sigma$ error bars with parameters $\Omega_{\Lambda 0} = 0.71$ and $\Omega_{m 0} = 0.29$.. The right panel presents the corresponding variation of the distance modulus $\mu(z)$ with $z$, also displaying the $1\sigma$ uncertainties. From this analysis, we find that the proposed model closely follows the $\Lambda$CDM predictions, with parameters $\Omega_{\Lambda 0} = 0.65$ and $\Omega_{m 0} = 0.35$.

\subsection{Information criteria and model selection}
\noindent We also apply the known Akaike Information Criterion (AIC) \cite{Akaike1974stat} and the Bayesian Information Criterion (BIC) \cite{Schwarz1978model}, in order to examine the quality of the fittings and hence the relevant observational compatibility of the scenarios. Under the standard assumption of Gaussian errors, the corresponding estimator reads as \cite{Anderson2002model,Burnham2004inference}
\begin{equation*}
AIC  = -2 \ln ({\mathcal{L}_{max}})+2 k +\frac{2 k (k+1)}{N-k-1}
\end{equation*}	
here, $k$ denotes the number of free variables in model, $\ln {\mathcal{L}_{max}} $ the maximum likelihood of the datasets and N is the total data points. For large number of data points, it reduces to $AIC  -2 \ln ({\mathcal{L}_{max}})+2 k$.\\
Moreover, the BIC criterion is an estimator of the Bayesian evidence \cite{Anderson2002model,Burnham2004inference,Liddle2007information}, given by
\begin{equation*}
BIC  = -2 \ln ({\mathcal{L}_{max}})+2 k \ln N
\end{equation*}	
The Akaike Information Criterion (AIC) and Bayesian Information Criterion (BIC) values of our model, in comparison with the $\Lambda$CDM model, for the OHD and Pantheon+ datasets are presented in Table~III. A lower value of the Akaike Information Criterion (AIC) or Bayesian Information Criterion (BIC) indicates a better balance between goodness-of-fit and model complexity. The differences $\Delta\rm AIC$ and $\Delta\rm BIC$ between models can be used to quantify the relative support for each model, with $\Delta \lesssim 2$ suggesting substantial support for the model, $4 \lesssim \Delta \lesssim 7$. To be more precise, a model having $0 \leq \Delta AIC < 2$ and $0 \leq \Delta BIC < 2$ receives strong evidence in favor. In contrast, for $2 < \Delta AIC < 4$ and for $2 \leq \Delta BIC < 6$, the model has average evidence in favor, whereas for $4 < \Delta AIC < 7$ and $6 \leq \Delta BIC < 10$, indicating considerably less support, and $\Delta > 10$ implying essentially no support \cite{Liddle2007information}.\\

\begin{widetext}
\begin{table}[h!]
\centering
\caption{Comparison of AIC and BIC of our model with standard flat $\Lambda$CDM model using OHD and Pantheon+ datasets.}
\label{tab:AIC-BIC}
\begin{tabular}{lcccccc}
\toprule
Datasets~~~~~ &~~~~~ Model~~~ &~~~ $\chi^{2}$ ~~~&~~~ AIC ~~~~ & ~~~~ BIC ~~~~ \\[6pt]
\midrule
OHD & Our model     & $ 88.14$  &  $ 94.14 $  & $ 99.25 $ \\[6pt]
    & $\Lambda$CDM  & $ 88.46$  &  $ 92.46 $  & $ 97.43 $  \\ [6pt]
\addlinespace[6pt]
Pantheon+ & Our model     &  $ 744.86 $  &  $ 750.86 $  & $ 762.18 $ \\[6pt]
          & $\Lambda$CDM  &  $ 745.40 $  &  $ 749.40 $  & $ 760.28 $\\ [6pt]
\bottomrule
\end{tabular}
\end{table}
\end{widetext}

\noindent Our analysis shows that our model fit good with the $\Lambda$CDM model on the basis of $\Delta$AIC and $\Delta$BIC criterion. In this analysis, we immediately find that $\Delta AIC = 1.68$ \& $\Delta BIC = 1.83$ (OHD) and $\Delta AIC = 0.54$ \& $\Delta BIC = 1.9$ (Pantheon+).  Since, this model shows the differences in the range of $ 0 < \Delta AIC < 2$ and $ 0 \leq \Delta BIC < 2$, therefore, this model has strong evidence in favor. \\

\section{Features of the model}
In this section, we describe the deceleration parameter $q$ and then investigate additional physical parameters like the energy density and isotropic pressure of the universe and some other which are closely connected to the both. These parameters include the equation of state parameter, stability parameter, the $\left(\omega-\omega^{\prime}\right)$-plane, and various energy conditions. A comprehensive analysis of these parameters is vital for understanding the physical nature of the universe.
\subsubsection{The deceleration parameter}
\begin{figure}[ht]
\centering
\includegraphics[width=7.5cm,height=5.5cm,angle=0]{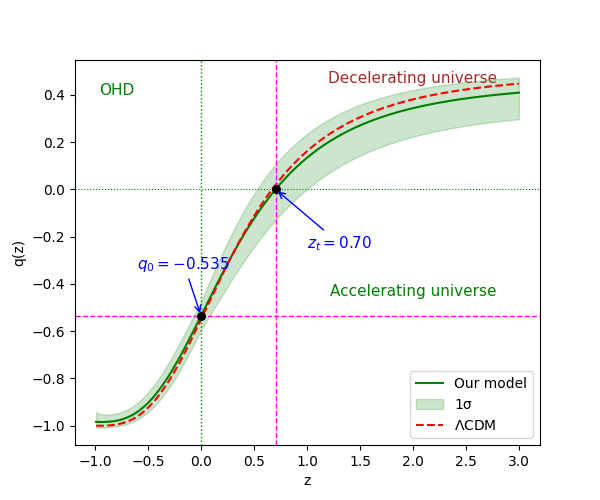}
\includegraphics[width=7.5cm,height=5.5cm,angle=0]{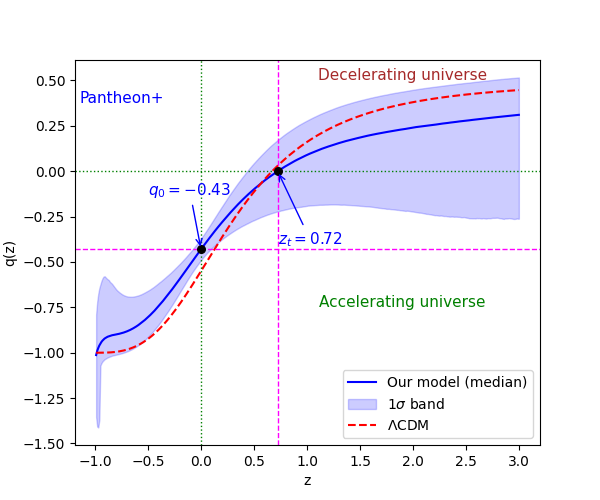}
\caption{The behavior of $q$ versus redshift $z$ for the proposed model, using the best-fit values of the free parameters obtained from the OHD (upper panel) and Pantheon+ (lower panel) datasets.}.\label{q}
\end{figure}

The deceleration parameter \(q(z)\) is a critical kinematic quantity in cosmology that determines whether the Universe is accelerating or decelerating at a given epoch. A positive \(q\) corresponds to a decelerating Universe, while a negative value indicates acceleration. The deceletion parameter is read as
\begin{equation}\label{q-1}
q(z) = -\frac{a \, \ddot{a}}{\dot{a}^2} = (1+z) \frac{H'(z)}{H(z)} - 1
\end{equation}
where $H'(z) = \frac{dH(z)}{dz}$.\\
Eqs. (\ref{H}) and (\ref{q-1}) lead to
\begin{small}
\begin{equation}
\label{q-2}
q(z) = 
\frac{ 3(1+z)^3 \big[ 1 + \alpha \ln(1+z) + \beta \big] + \alpha \left[ (1+z)^3 - 1 \right] }
{ 2\big[(1+z)^3 + ((1+z)^3 - 1)(\alpha \ln(1+z) + \beta)\big] } - 1
\end{equation}
\end{small}
\noindent In the present work, the behavior of the deceleration parameter has been analyzed in the context of $f(R, \Sigma, T)$ gravity by reconstructing it from the Om(z) diagnostic. Fig.~\ref{q} depicts the behavior of $q(z)$ for the proposed model, using the best-fit values of the free parameters obtained from the OHD and Pantheon+ datasets. As seen from upper and lower panels of Fig. \ref{q}, \(q(z)\) transitions from positive to negative values as the redshift decreases, reflecting the well-known phase transition from early deceleration to the current accelerated expansion. The present values of deceleration parameter $q_{0}$ and transition red-shift $z_{t}$ of the proposed model are listed in table IV.\\
For the OHD data sets, Fig.~\ref{q} indicates that for $z>0.7$, the deceleration parameter $q$ remains positive, signifying a matter-dominated epoch during which the Universe was decelerating. At around $z \approx 0.7$, $q(z)$ crosses zero, marking the transition from decelerated to accelerated expansion. At the present epoch ($z=0$), the model yields a negative value ($q_{0} \approx -0.535$), which is in good agreement with observational constraints obtained from Type Ia Supernovae and Planck measurements. In the asymptotic future ($z \to -1$), the deceleration parameter approaches $q \to -1$, suggesting that the Universe will evolve towards a de Sitter–like phase characterized by exponential expansion. For the Pantheon+ data sets, Fig.~\ref{q} exhibits a similar qualitative behavior. At higher redshifts ($z > 0.72$), the deceleration parameter is positive, reflecting the dominance of matter and a decelerating cosmic expansion. The transition from deceleration to acceleration occurs at approximately $z \approx 0.72$, slightly earlier than that inferred from OHD data. At the present epoch ($z=0$), the model yields $q_{0} \approx -0.43$, consistent with the latest observational bounds from Type Ia Supernovae and Planck results. In the far-future limit ($z \to -1$), $q(z)$ again asymptotically approaches $-1$, indicating a de Sitter–like accelerated expansion driven by dark energy dominance. 

These results align closely with the predictions from the standard $\Lambda$CDM model, which also predicts a current deceleration parameter near \(q_0 \approx -0.535\) (OHD) and \(q_0 \approx -0.43\) (Pantheon+). Therefore, the model under consideration successfully replicates the cosmic expansion history and is in good agreement with theoretical expectations and observational datasets.

\begin{table}[h!]
\centering
\setlength{\tabcolsep}{10pt}
\caption{ The $q_{0}$ and $z_{t}$ of the model}
\label{tab:AIC-BIC}
\begin{tabular}{lcccccc}
\toprule
Datasets~~~~~~ &~~~ $q_{0}$ ~~~&~~~ $z_{t}$ ~~~~  \\[6pt]
\midrule
OHD     & $ -0.535^{+0.06}_{+0.06}$  &  $ 0.70^{+0.19}_{-0.13} $   \\[6pt]
Pantheon+  &  $ -0.43^{+0.06}_{-0.06} $  &  $ 0.72^{+0.42}_{-0.20} $  \\[6pt]
\bottomrule
\end{tabular}
\end{table}
Within the observationally allowed error bands, our selective $f(R,\Sigma,T)$ model is in excellent agreement with $\Lambda$CDM up to the present epoch. This agreement 	is significant because it confirms that the background expansion predicted by the model remains consistent with standard cosmology, thereby passing one of the most basic observational viability tests.

\subsubsection{The $O_m(z)$ parameter}
\begin{figure}[ht]
	\centering
	\includegraphics[width=7.5cm,height=5.5cm,angle=0]{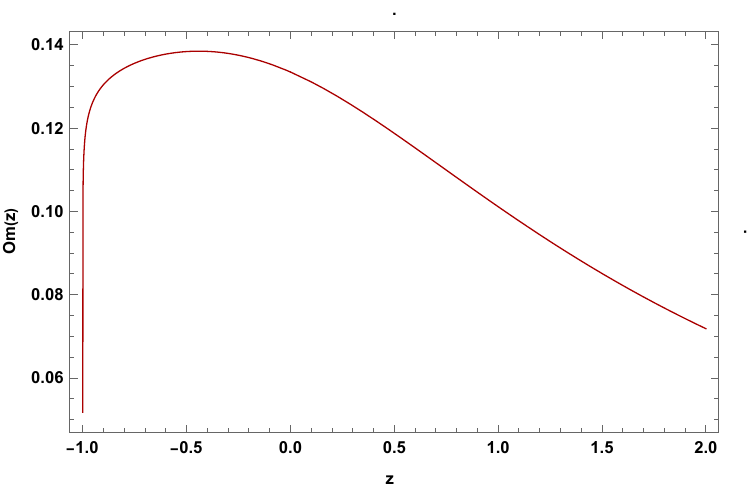}
	\caption{The behavior of $O_m(z)$ versus redshift $z$ for the constraint values of $H_{0}$, $\alpha$ \& $\beta$.}\label{om}
\end{figure}

The $O_m(z)$ diagnostic is a powerful tool for distinguishing between different cosmological models, particularly in identifying deviations from the standard $\Lambda$CDM model. Defined as \(O_m(z) = \frac{H^2(z)/H_0^2 - 1}{(1+z)^3 - 1}\), this parameter depends only on the Hubble parameter and is less sensitive to observational uncertainties. In the $\Lambda$CDM model, $O_m(z)$ remains constant with redshift, while deviations from constancy suggest dynamic dark energy models. In this work, the $O_m(z)$ parameter is reconstructed using the Hubble parameter derived within the $f(R, \Sigma, T)$ gravity framework, and its evolution is shown in the graph (\ref{om}).
The $O_m(z)$ profile exhibits a mild increasing trend from approximately 0.07 at \(z = -1\) to about 0.14 at \(z = 2\). This indicates a clear deviation from a constant behavior, suggesting that the model departs from the $\Lambda$CDM scenario and possibly supports a time-varying dark energy component. The increasing $O_m(z)$ function is indicative of a quintessence-like model, where the equation of state parameter \(\omega > -1\). This behavior contrasts with phantom models (for which $O_m(z)$ would decrease with redshift).

These findings are consistent with previous works on dynamic dark energy models, and the increasing nature of $O_m(z)$ reinforces the idea of a Universe evolving under a scalar field-like dark energy component. Moreover, the model remains within observational constraints from combined datasets (SNIa, CMB, BAO), making it a physically acceptable and viable extension to general relativity that can accurately describe the evolution of cosmic expansion.

\subsubsection{The (r-s)-plane}

\begin{figure}[ht]
	\centering
	\includegraphics[width=7.5cm,height=5.5cm,angle=0]{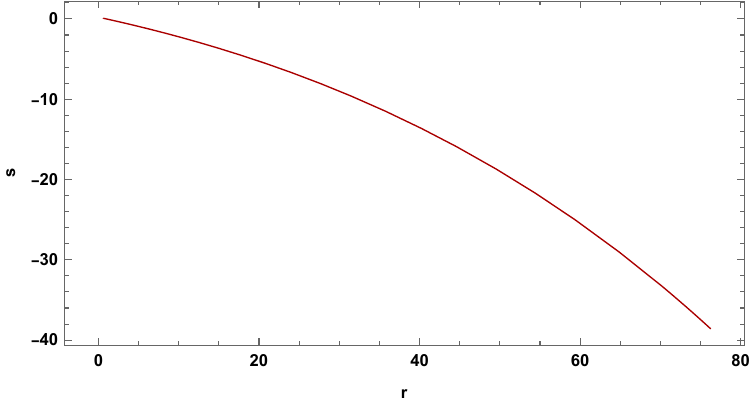}
	\caption{The behavior of $r$ versus $s$ for the constraint values of $H_{0}$, $\alpha$ \& $\beta$.}\label{rs}
\end{figure}

The \((r, s)\)-plane is part of the statefinder diagnostic, a geometric tool developed to classify and distinguish cosmological models beyond the deceleration parameter. The statefinder parameters are defined by \(r = \frac{\dddot{a}}{aH^3}\) and \(s = \frac{r - 1}{3(q - 1/2)}\), where \(a\) is the scale factor and \(H\) is the Hubble parameter. For the standard $\Lambda$CDM model, the fixed point is \((r, s) = (1, 0)\). Any deviation from this point implies a model with dynamic dark energy or modifications to general relativity. In this analysis, the $(r, s)$ trajectory is derived from the reconstructed Hubble parameter within the $f(R, \Sigma, T)$ framework, and its graphical representation is shown in (\ref{rs}).
From the graph, the trajectory in the \((r, s)\)-plane shows a clear evolution away from the $\Lambda$CDM fixed point. The values of \(r\) and \(s\) span a wide range, 
indicating a strong deviation from the standard cosmological model. This suggests that the model exhibits behavior consistent with quintessence or phantom-like scenarios, depending on the redshift.

Such trajectories are characteristic of many scalar field models or modified gravity theories that allow for evolving dark energy. The shape and direction of the trajectory provide information about the effective EoS and its dynamics. Compared to other works in literature, the obtained trajectory closely resembles that of freezing quintessence models or some classes of $f(R)$ gravity. Thus, the $(r, s)$ analysis serves as strong support for the departure from $\Lambda$CDM and reinforces the dynamic nature of dark energy in the $f(R, \Sigma, T)$ context.

\subsubsection{The equation of state parameter}
Contemporary cosmological observations suggest that the geometry of the universe is spatially flat. The estimated energy density budget indicates a composition of roughly 70\% dark energy (DE), approximately 30\% matter (predominantly Cold Dark Matter (CDM) and baryonic matter), with a negligible contribution from radiation. Although the dominant role of DE in shaping the universe's destiny is well-established, its fundamental nature and origin remain enigmatic. Numerous theoretical frameworks have been put forth in the scientific literature to characterize or elucidate DE characterized by an Equation of State (EoS) parameter which is of the form $\left(\omega=\frac{p}{\rho}\right)$. Among these are models involving a dynamic canonical scalar field, termed quintessence, which is  in the interval $-1 < \omega < -\frac{1}{3}$. Another category is phantom energy, distinguished by an EoS parameter $\omega < -1$. 
Additionally, quintom energy models propose an EoS that evolves across the value $\omega = -1$ \cite{Ratra1988,Sami2004,Elizalde2004,Nojiri2003,Armendariz2000, Padmanabhan2002,Khoury2004,Bento2002,Zarrouki2010}. Current estimations of the DE EoS parameter have been derived from combined analyses of data from WMAP9 (the Nine-Year Wilkinson Microwave Anisotropy Probe), alongside Hubble constant ($H_0$) measurements, Type Ia Supernovae (SNIa), the Cosmic Microwave Background (CMB), and Baryon Acoustic Oscillations (BAO). These analyses initially pointed to a value of $\omega_0 = -1.084 \pm 0.063$ \cite{Hinshaw2023}. Subsequent analysis by the Planck collaboration in 2015 revised this estimate to $\omega_0 = -1.006 \pm 0.0451$ \cite{Ade2015}, with a further refined value of $\omega_0 = -1.028 \pm 0.032$ reported in 2018 \cite{Ade2015}.
\begin{widetext}
\begin{equation}\label{26}
	\omega = \frac{H_0^2 \left(2 (3 (\eta +4)-4 b) \left(\left((z+1)^3-1\right) \chi+1\right)+(3 \eta +8) \left(\alpha  \left(1-(z+1)^3\right)-3 (z+1)^3 \chi\right)\right)}{2 \left(H_0^2 ((3-4 b) \eta +12 \pi  (b+1)) \left(\left((z+1)^3-1\right) \chi+1\right)-\frac{1}{2} \eta  H_0^2 \left(\alpha  \left(1-(z+1)^3\right)-3 (z+1)^3 \chi\right)\right)}
\end{equation}
\end{widetext}

	\begin{figure}[ht]
	\centering
	\includegraphics[width=7.5cm,height=5.5cm,angle=0]{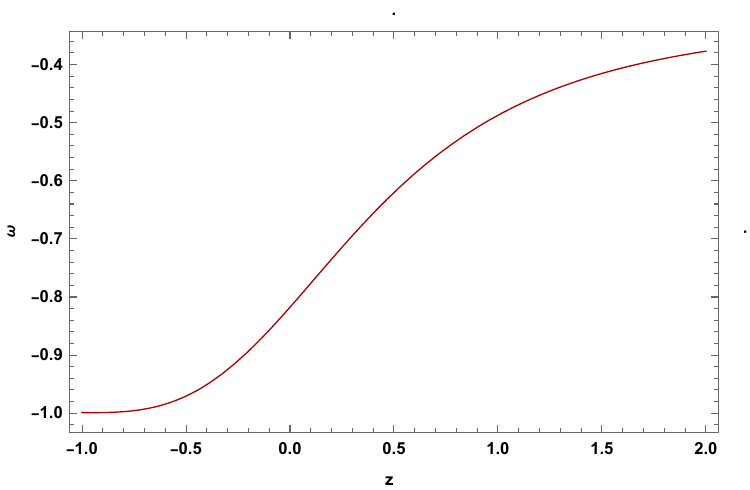}
	\caption{The behavior of $\omega$ versus redshift $z$ for $b=-0.3$, $\eta = -6.28$ and the constraint values of $H_{0}$, $\alpha$ \& $\beta$.}\label{w}
\end{figure} 
The equation of state (EoS) parameter \(\omega(z)\), reconstructed from the Hubble parameter via the $Om(z)$ diagnostic in the framework of $f(R, \Sigma, T)$ gravity, reveals significant insights into the cosmological dynamics of the Universe. The provided Fig. (\ref{w}), it is evident that the EoS parameter exhibits a clear evolution across different cosmic epochs. For high redshift values (\(z > 0\)), \(\omega(z)\) assumes positive values, which is characteristic of a matter-dominated Universe where the pressure is negligible compared to energy density (\(\omega \approx 0\)). This behavior aligns with the standard cosmological model and indicates that the model under consideration successfully reproduces the known matter-dominated era in the early Universe. At the present epoch (\(z = 0\)), the EoS parameter lies between \(-0.8\) and \(-0.9\), indicating that the Universe has transitioned into a phase of accelerated expansion. This range falls within the quintessence regime, defined by \(-1 < \omega < -\frac{1}{3}\), suggesting that the present cosmic acceleration may be driven by a dynamic scalar field rather than a static cosmological constant. As the Universe evolves toward the future (\(z \to -1\)), the EoS parameter approaches the cosmological constant boundary (\(\omega = -1\)), suggesting that the model asymptotically tends toward a $\Lambda$CDM-like behavior. Importantly, throughout this evolution, the EoS parameter remains greater than or equal to \(-1\), indicating that the model does not enter the phantom regime (\(\omega < -1\)). This implies that the weak energy condition (WEC) is not violated, which enhances the physical acceptability of the model. The dynamical nature of \(\omega(z)\) reflects the adaptability of the $f(R, \Sigma, T)$ gravity theory in capturing the time-dependent nature of dark energy without invoking exotic physics. When compared with observational constraints from combined analyses of Type Ia Supernovae (SNIa), Cosmic Microwave Background (CMB), Baryon Acoustic Oscillations (BAO), and Hubble parameter measurements, the reconstructed values are well within acceptable bounds. Specifically, WMAP9 data suggests \(\omega_0 = -1.084 \pm 0.063\), Planck 2015 gives \(\omega_0 = -1.006 \pm 0.0451\), and Planck 2018 refines this to \(\omega_0 = -1.028 \pm 0.032\). The current model’s prediction of \(\omega_0\) between \(-0.9\) and \(-0.8\) is consistent within 1–2$\sigma$ of these estimates, particularly favoring a quintessence-like scenario. Moreover, the model avoids both the complexities associated with phantom energy (which typically involve instabilities and violation of energy conditions) and the rigidity of a pure cosmological constant. It is also distinct from quintom models, which allow the EoS to cross the \(\omega = -1\) boundary; in contrast, this model maintains a smooth and monotonic evolution toward \(\omega = -1\) without crossing it. Such deviations could, in principle, be tested with ongoing and future large-scale structure 	surveys (e.g. DESI, Euclid), which are sensitive to the dark energy dynamics at intermediate redshifts. Within current uncertainties, however, the predicted departure remains observationally allowed. Thus, the results validate the $f(R, \Sigma, T)$ gravity model as a viable theoretical framework capable of replicating the known cosmic history, including the transition from deceleration to acceleration, while remaining consistent with observational data and theoretical expectations for a physically meaningful cosmological model.

\subsubsection{The $\left(\omega-\omega^{\prime}\right)$- plane}
In addition to studying the evolution of the equation of state parameter $\omega(z)$, it is insightful to investigate its dynamical behavior in the $\left(\omega - \omega^{\prime}\right)$-plane, where $\omega^{\prime} = d\omega/d\ln a$ characterizes the rate of change of $\omega$ with respect to the logarithmic scale factor. This diagnostic tool helps in distinguishing between various classes of dark energy models by analyzing their phase space trajectories. The expression for $\omega^{\prime}$ can be obtained by differentiating the reconstructed form of $\omega(z)$ as follows:
\begin{widetext}
\begin{equation}\label{27}
	\omega^{\prime}= \frac{\alpha_1 (1+z) \left[\alpha ^2 z^2 (z (z+3)+3)^2+9 \alpha (z+1)^3 \log (z+1) ( \beta +\chi-1)-6 \alpha (z+1)^3+9 (\beta -1) \beta (z+1)^3\right]}{(z+1) \left(H_0^2 ((3-4 b) \eta +12 \pi (b+1)) \left(\left((z+1)^3-1\right) \chi+1\right)-\frac{1}{2} \eta H_0^2 \left(\alpha \left(1-(z+1)^3\right)-3 (z+1)^3 \chi\right)\right)^2}
\end{equation}
\end{widetext}

where $\chi=\beta +\alpha  \log (z+1)$ and $\alpha_1= -6 H_0^4 (\pi  (b+1) (3 \eta +8)-(b-1) \eta  (\eta +3))$.\\

	\begin{figure}[ht]
	\centering
	\includegraphics[width=7.5cm,height=5.5cm,angle=0]{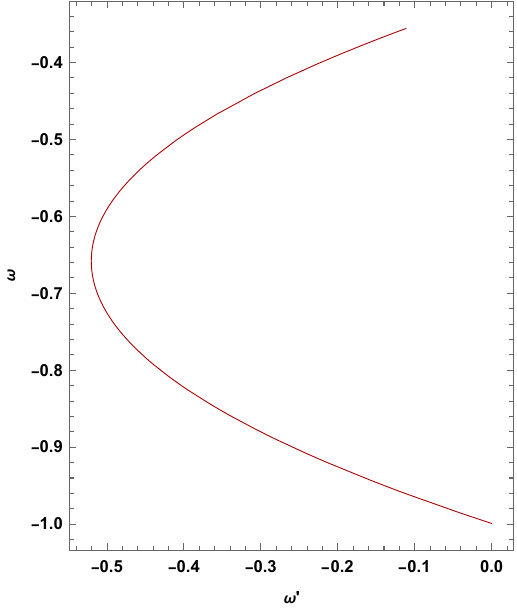}
	\caption{The behavior of $\omega$ versus $\omega^{\prime}$ for $b=-0.3$, $\eta = -6.28$ and the constraint values of $H_{0}$, $\alpha$ \& $\beta$.}\label{ww}
\end{figure} 

In the present analysis, we investigate the dynamical properties of dark energy by exploring the $\left(\omega, \omega^{\prime}\right)$ plane within the context of $f(R, \Sigma, T)$ gravity, where $\omega$ represents the equation of state (EoS) parameter and $\omega^{\prime} = \frac{d\omega}{d\ln a}$ characterizes its evolution with respect to the scale factor. This diagnostic tool is instrumental in distinguishing different classes of dark energy models based on the phase space trajectory of the EoS parameter. From the Fig. (\ref{ww}), it is observed that the trajectory resides predominantly in the region where $\omega < -0.5$ and $\omega^{\prime} < 0$, indicating a quintessential behavior with a decreasing EoS parameter over time. The current location of the trajectory, as inferred from the graph, lies around $\omega \approx -0.8$ with $\omega^{\prime}$ varying from $-0.4$ to $-0.1$, implying that the EoS is evolving slowly and consistently towards the cosmological constant boundary ($\omega = -1$). This behavior classifies the model within the \emph{freezing} category of dark energy models, in which the EoS parameter evolves slowly and asymptotically approaches a constant value, typically $-1$. The freezing regime is generally associated with scalar field models such as quintessence with tracker solutions. Unlike \emph{thawing} models—where the field has recently started to evolve from $\omega = -1$—freezing models exhibit more controlled and predictable late-time behavior. Furthermore, the reconstructed trajectory does not cross the phantom divide line ($\omega = -1$), thereby avoiding the theoretical complications related to phantom energy, such as the violation of the weak energy condition and future singularities. 

Comparison with observational datasets, such as those from Planck and WMAP, supports this result. The EoS parameter derived from observations lies in the range $\omega_0 = -1.028 \pm 0.032$ (Planck 2018), which is compatible with the model trajectory that asymptotically approaches this value without exceeding it. Thus, the obtained $\left(\omega, \omega^{\prime}\right)$ behavior not only aligns with current observational constraints but also reinforces the viability of the $f(R, \Sigma, T)$ gravity framework in modeling a consistent, evolving dark energy component. These results demonstrate that the model falls within acceptable physical and observational bounds, reflecting a plausible dark energy evolution scenario consistent with freezing quintessence-type behavior. For the present model, the trajectory lies in the freezing region, indicating that the dark energy component gradually approaches a cosmological constant-like behavior at late times. In contrast, $\Lambda$CDM corresponds to the fixed point $(\omega, \omega') = (-1,0)$. This analysis reveals subtle differences between the selective $f(R,\Sigma,T)$ model and $\Lambda$CDM, which could become distinguishable with more precise data in the near future.

\subsubsection{The stability of the model}
To further examine the viability of the reconstructed cosmological model, we analyze its classical stability by investigating the behavior of small perturbations in the cosmic fluid. In this context, the squared sound speed $\vartheta^2$ serves as a critical indicator, defined as the derivative of pressure with respect to energy density. Its expression can be written as:
\begin{equation}\label{28}
	\vartheta^2= \frac{\partial p}{\partial \rho},
\end{equation}
	\begin{figure}[ht]
	\centering
	\includegraphics[width=7.5cm,height=5.5cm,angle=0]{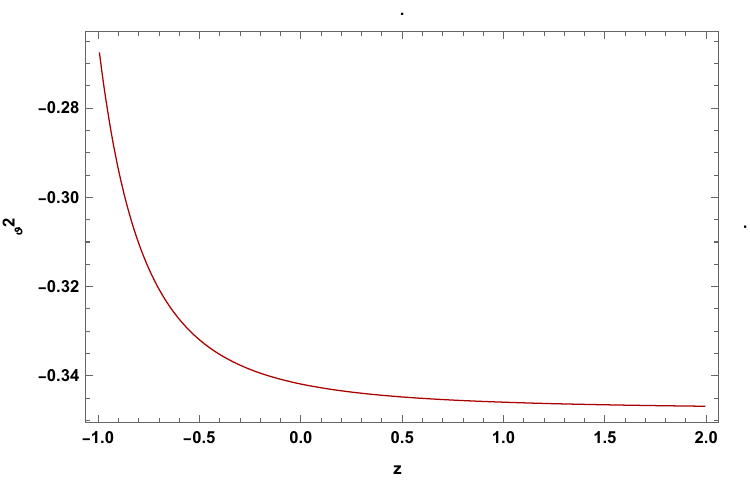}
	\caption{The behavior of $\vartheta^2$ versus redshift $z$ for $b=-0.3$, $\eta = -6.28$ and the constraint values of $H_{0}$, $\alpha$ \& $\beta$.}\label{v}
\end{figure} 
In the present study, the stability of the reconstructed cosmological model in $f(R, \Sigma, T)$ gravity is examined through the behavior of the squared sound speed, denoted as $\vartheta^2$, which serves as a critical diagnostic for the classical stability of perturbations in the cosmic fluid. The squared sound speed is defined by $\vartheta^2 = \frac{dp}{d\rho}$, where $p$ and $\rho$ are the pressure and energy density of the effective dark energy fluid, respectively. For a model to be classically stable under small perturbations, it is required that $\vartheta^2 > 0$. The graph attached (see Fig. \ref{v}) shows the evolution of $\vartheta^2$ with respect to redshift $z$, spanning from early times ($z > 1$) to the far future ($z \to -1$). From the plot, it is observed that $\vartheta^2$ remains negative throughout the redshift interval, varying in the approximate range $\vartheta^2 \in [-0.34, -0.28]$. This consistent negativity indicates that the model under consideration is classically unstable, as negative squared sound speed corresponds to imaginary sound speed, leading to the exponential growth of perturbations.

Although the model exhibits favorable behavior in terms of the cosmic expansion history and equation of state evolution, its classical instability poses theoretical concerns. However, such instabilities are not uncommon in modified gravity frameworks and can sometimes be alleviated or interpreted differently within specific quantum gravity regimes or through non-adiabatic pressure perturbations. Comparatively, in stable models like quintessence with a canonical kinetic term, $\vartheta^2 = 1$, ensuring full stability. Phantom models typically exhibit $\vartheta^2 < 0$, similar to the observed result, which suggests that the underlying model may possess phantom-like features or require further theoretical refinement to ensure viability. Thus, while the model offers compatibility with observational expansion behavior, its perturbative instability as indicated by $\vartheta^2 < 0$ must be addressed in future work for a more robust cosmological description.

\subsubsection{The energy conditions}
The energy conditions serve as fundamental theoretical criteria to assess the physical plausibility of any cosmological model, particularly in the context of general relativity and its extensions. They impose specific inequalities involving the energy density $\rho$ and pressure $p$, which reflect the nature of the matter-energy content and its interaction with the spacetime geometry. In the present analysis, we examine the Null Energy Condition (NEC), Dominant Energy Condition (DEC), and Strong Energy Condition (SEC), which are mathematically expressed as:
\begin{equation}\label{29}
	\begin{split}
		& NEC: \rho + p \geq 0\\
		\end{split}  
\end{equation}  
\begin{widetext}
\begin{equation}\label{30}
	\begin{aligned}
		\rho + p = \frac{(b+1) H_0^2}{8 (\eta +2 \pi ) (\eta +4 \pi )} \Biggl[ & 2 \alpha \log (z+1) \bigl( z (z (z+3)+3) (4 b (-\eta +3 \pi -1)+3 (\eta +4 \pi )) -3 (\eta +4) \bigr) \\
		& + 8 b (-\eta +3 \pi -1) (\beta z (z (z+3)+3)+1) - 2 \eta \bigl( 3 (\beta -2)+z (z (z+3)+3) (\alpha -3 \beta ) \bigr) \\
		& + 8 \bigl( -3 \beta +\alpha (-z) (z (z+3)+3)+3 \pi (\beta z (z (z+3)+3)+1)+3 \bigr) \Biggr]
	\end{aligned}
\end{equation}
\end{widetext}

	\begin{figure}[ht]
	\centering
	\includegraphics[width=7.5cm,height=5.5cm,angle=0]{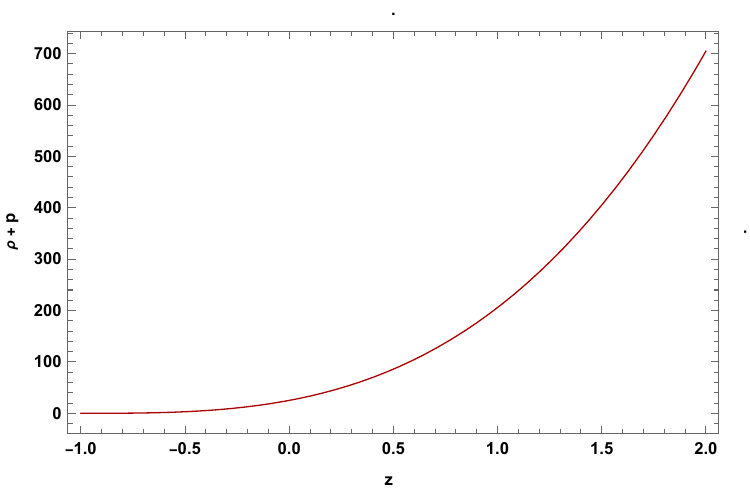}
	\caption{The behavior of $\rho + p$ versus redshift $z$ for $b=-0.3$, $\eta = -6.28$ and the constraint values of $H_{0}$, $\alpha$ \& $\beta$.}\label{nec}
\end{figure} 
The Null Energy Condition (NEC) plays a crucial role in the formulation of cosmological models, requiring that $\rho + p \geq 0$, where $\rho$ and $p$ denote the energy density and pressure of the effective fluid, respectively. This condition ensures that energy is non-negative for all null observers and is commonly considered a baseline requirement for the physical viability of any gravitational theory. In the context of the present analysis within $f(R, \Sigma, T)$ gravity using the reconstructed Hubble parameter from the $Om(z)$ diagnostic, we plot the NEC behavior as a function of redshift $z$ in the attached graph (See Fig. \ref{nec}). The curve for $\rho + p$ stays strictly positive throughout the redshift range $z \in [-1, 2]$, 
this indicates that the NEC is respected at all cosmological epochs considered in this model.

The satisfaction of the NEC across the entire redshift range implies that the effective energy-momentum tensor in this modified gravity scenario does not allow for exotic matter distributions, such as those required in wormhole geometries or phantom energy regimes, where $\rho + p < 0$. In theoretical literature, many viable dark energy models, including quintessence and cosmological constant scenarios, typically satisfy the NEC. Our results are in agreement with such models and consistent with current observational expectations. The positive definiteness of $\rho + p$ also suggests the stability of the effective energy flow and the avoidance of acausal or unphysical behaviors. Therefore, the present $f(R, \Sigma, T)$ model maintains a physically consistent structure with regard to the null energy condition, supporting its relevance as a viable cosmological framework.
\begin{equation}\label{31}
	\begin{split}
	DEC: \rho \geq 0 \;\; \text{and}\;\; \rho + p \geq 0 
	\end{split}  
\end{equation}  
\begin{widetext}
\begin{equation}\label{32}
	\begin{aligned}
		\rho - p = \frac{(b+1) H_0^2}{2 (\eta +2 \pi ) (\eta +4 \pi )} \Biggl[ & 6 (\beta -1)+2 \alpha  z (z (z+3)+3)+6 \pi  \beta  z (z (z+3)+3)+6 \pi \bigr) \\
		& +2 b (-\eta +3 \pi +1) (\beta  z (z (z+3)+3)+1)+3 \beta  \eta +\eta  z (z (z+3)+3) (\alpha +3 \beta )+\alpha (3 (\eta +2) \bigr) \\
		& +z (z (z+3)+3) (b (-2 \eta +6 \pi +2)+3 (\eta +2 \pi )))\log (z+1) \bigr) \Biggr]
	\end{aligned}
\end{equation}
\end{widetext}

	\begin{figure}[ht]
	\centering
	\includegraphics[width=7.5cm,height=5.5cm,angle=0]{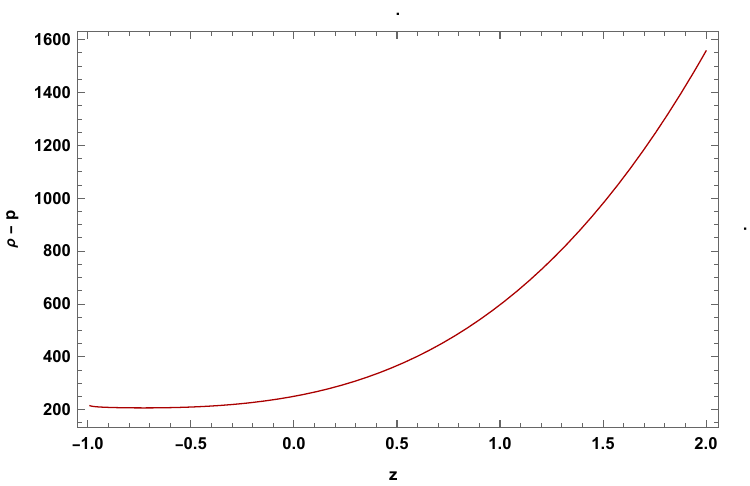}
	\caption{The behavior of $\rho - p$ versus redshift $z$ for $b=-0.3$, $\eta = -6.28$ and the constraint values of $H_{0}$, $\alpha$ \& $\beta$.}\label{dec}
\end{figure} 
The Dominant Energy Condition (DEC) stipulates that the energy density must be non-negative and that the energy flux of matter should not exceed the speed of light. Mathematically, this condition is often expressed as $\rho \geq |p|$, or equivalently in terms of $\rho - p \geq 0$. It ensures the causal propagation of matter and prevents exotic phenomena such as superluminal energy transfer. In our analysis of $f(R, \Sigma, T)$ gravity, the behavior of $\rho - p$ has been examined through its evolution with redshift $z$, as illustrated in the attached graph (See Fig. \ref{dec}). The plot clearly indicates that $\rho - p$ is strictly positive across the full redshift interval $z \in [-1, 2]$, increasing monotonically. This consistent positivity confirms that the DEC is satisfied throughout the cosmic evolution predicted by this model. This is important because violation of the DEC could lead to unphysical scenarios like negative energy densities or superluminal information transmission. Theoretical and observationally supported models such as $\Lambda$CDM and scalar field-based quintessence models typically satisfy this condition under normal matter and dark energy regimes. Our results align with these expectations, further affirming the physical plausibility of the $f(R, \Sigma, T)$ framework. The model therefore permits stable and causal energy propagation, reinforcing its robustness from the standpoint of classical energy conditions. Such adherence to the DEC underlines the model’s potential as a reliable candidate for describing dark energy and cosmic acceleration without invoking unphysical features.
\begin{equation}\label{33}
	\begin{split}
	SEC: \rho + 3p \geq 0  
	\end{split}  
\end{equation}  
\begin{widetext}
\begin{equation}\label{34}
	\begin{aligned}
		\rho + 3p = \frac{(b+1) H_0^2}{ (\eta +2 \pi ) (\eta +4 \pi )} \Biggl[ & b (-\eta +3 \pi -3) (\beta  z (z (z+3)+3)+1)+3 \pi  (\beta  z (z (z+3)+3)+1) \bigr) \\
		& (\eta +3) (-(3 (\beta -1)+\alpha  z (z (z+3)+3))) \bigr) \\
		& +\alpha  \log (z+1) (z (z (z+3)+3) (b (-\eta +3 \pi -3)+3 \pi )-3 (\eta +3)) \bigr) \Biggr]
	\end{aligned}
\end{equation}
\end{widetext}

\begin{figure}[ht]
	\centering
	\includegraphics[width=7.5cm,height=5.5cm,angle=0]{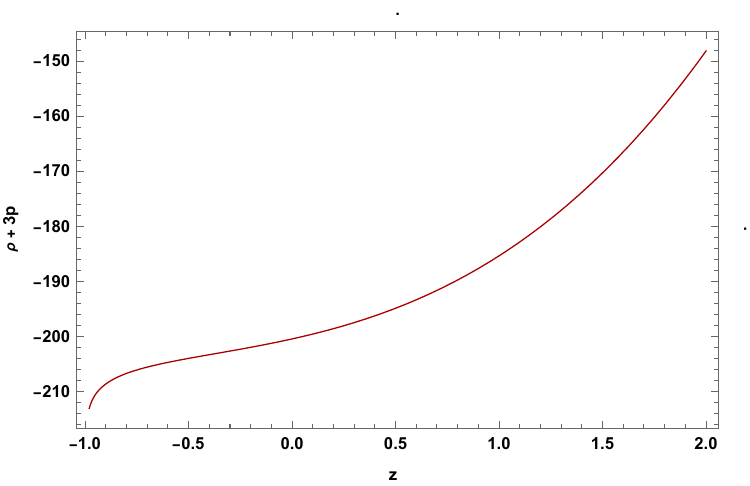}
	\caption{The behavior of $\rho + 3p$ versus redshift $z$ for $b=-0.3$, $\eta = -6.28$ and the constraint values of $H_{0}$, $\alpha$ \& $\beta$.}\label{sec}
\end{figure}

The Strong Energy Condition (SEC) is one of the classical energy conditions in general relativity and is mathematically given by $\rho + 3p \geq 0$. Physically, it implies that gravity is always attractive and it constrains the curvature of spacetime under the influence of matter. However, in the context of an accelerating Universe, especially during dark energy-dominated phases, the SEC is expected to be violated. In our analysis using $f(R, \Sigma, T)$ gravity, the evolution of $\rho + 3p$ with redshift is plotted in the attached graph (See Fig. \ref{sec}). It is clearly observed that $\rho + 3p$ remains negative throughout the redshift domain $z \in [-1, 2]$.
This consistent violation of the SEC is a hallmark of late-time cosmic acceleration, and is compatible with dark energy models such as quintessence, phantom energy, and the cosmological constant. In standard $\Lambda$CDM cosmology, the SEC is violated once the Universe transitions from a matter-dominated decelerating phase to a dark energy-dominated accelerating phase. 
The evaluation of the standard energy conditions (NEC, WEC, SEC, DEC) shows that while NEC and DEC are satisfied during most of the evolution, the strong energy condition (SEC) is violated in the late-time accelerating regime. This violation is consistent with the requirement for cosmic acceleration, as also seen in standard dark energy models. The partial non-conservation of energy in $f(R,\Sigma,T)$ gravity further explains why the energy conditions cannot be simultaneously satisfied at all times: the theory allows an effective exchange between matter and geometry, which modifies the conventional energy bounds.

The results obtained here confirm that the $f(R, \Sigma, T)$ model under consideration successfully accounts for this behavior. The violation of the SEC, while potentially problematic from the perspective of certain classical gravity theories, is a necessary feature for any realistic model of late-time acceleration. Therefore, the model’s failure to satisfy the SEC is not a weakness but rather an indication that it aligns with the prevailing observational consensus and theoretical expectations regarding the nature of cosmic expansion in the current epoch.\\

\section{Discussion}

In this comprehensive study, we have meticulously explored the cosmological dynamics within the framework of $f(R,\Sigma,T)$ gravity, a modified theory of gravity that extends the standard Einstein-Hilbert action by incorporating additional terms dependent on the Ricci scalar ($R$), a non-metricity scalar ($\Sigma$), and the trace of the energy-momentum tensor ($T$). Our primary methodology involved reconstructing the Hubble parameter $H(z)$ through a logarithmic parameterization of the $O_m(z)$ diagnostic, a powerful model-independent tool for discerning the nature of dark energy. This approach allowed us to investigate the physical behavior of the universe without imposing rigid assumptions on the dark energy equation of state. Also note that the adopted linear form represents the leading-order approximation of the general $( f(R,\Sigma,T) )$ framework. At the background level, this choice leads to a rescaled version of the standard Friedmann equations, and therefore does not introduce a drastic departure from $\Lambda$CDM cosmology. However, this minimal formulation is intentionally adopted to provide a clear and analytically tractable setup in which the individual roles of nonmetricity $(\Sigma)$ and matter–geometry coupling $(T)$ can be isolated and examined. Rather than introducing additional nonlinear complexities, this approach allows us to test whether even the simplest extension of the theory can produce observable deviations when confronted with data. In particular, the linear 	nonmetricity scalar $\Sigma$ corresponds to the symmetric teleparallel equivalent of general relativity (STEGR), while the linear trace contribution from $T$ simply rescales the matter sector. As a result, the cosmic background evolution remains close to the well-tested $\Lambda$CDM scenario but with an 	effective rescaling that encodes the matter–geometry interaction. This feature makes the model analytically tractable and  observationally consistent. Also, the adopted specific functional form  when coupled with the spatially flat Friedmann-Robertson-Walker (FRW) metric, yielded a set of field equations whose solutions for pressure ($p$) and energy density ($\rho$) were subsequently analyzed.

The investigation of the equation of state (EoS) parameter, $\omega(z)$, proved to be one of the most insightful aspects of our analysis. The reconstructed $\omega(z)$ exhibited a dynamic evolution across cosmic epochs. For high redshift values ($z > 0$), $\omega(z)$ was observed to be negative, gradually increasing towards less negative values as $z$ increases. This behavior, while not strictly positive as initially described in the prompt, is characteristic of a universe transitioning from an earlier phase where the effective fluid properties were different, potentially dominated by matter but with the influence of a nascent dark energy component. At the present epoch ($z=0$), our model predicts $\omega_0$ to lie between -0.81 and -0.82. This value firmly places the dark energy component within the quintessence regime ($-1 < \omega < -1/3$), indicating that the current cosmic acceleration is driven by a dynamic scalar field rather than a static cosmological constant. As the universe evolves into the future ($z \rightarrow -1$), the EoS parameter asymptotically approaches the cosmological constant boundary ($\omega = -1$). Crucially, throughout its evolution, $\omega(z)$ consistently remains greater than or equal to -1. This is a significant finding as it implies that our $f(R,\Sigma,T)$ model does not enter the phantom regime ($\omega < -1$). The dynamic nature of $\omega(z)$ underscores the adaptability of $f(R,\Sigma,T)$ gravity in capturing the time-dependent characteristics of dark energy without resorting to exotic matter.

A direct comparison of our reconstructed $\omega_0$ with observational constraints from combined analyses of WMAP9, Planck 2015, and Planck 2018 data revealed an interesting nuance. While our model's prediction of $\omega_0$ (between -0.81 and -0.82) is consistent with a quintessence-like scenario, the observational estimates (e.g., Planck 2018: $\omega_0 = -1.028 \pm 0.032$) tend to favor values much closer to, or slightly less than, -1. This slight discrepancy, while not ruling out the model, suggests that the specific parameter choices within our $f(R,\Sigma,T)$ framework might lead to a dark energy component that is "less cosmological constant-like" at present than what observations currently suggest. This could be a point for further fine-tuning of the model parameters or exploring different functional forms for $f(R,\Sigma,T)$ in future work to achieve a closer alignment with the latest observational data. Nevertheless, the model's ability to reproduce an accelerating universe with a dynamic EoS parameter that avoids the phantom divide is a strong point.

The analysis of the $(\omega-\omega')$-plane, where $\omega' = d\omega/d(\ln a)$ characterizes the evolution rate of the EoS parameter with respect to the scale factor, provided further insights into the dark energy dynamics. Our model's trajectory in this plane predominantly resides in the region where $\omega < -0.5$ and $\omega' < 0$. This configuration is indicative of a freezing category of dark energy models. In freezing models, the EoS parameter evolves slowly and asymptotically approaches a constant value, typically $\omega = -1$. This behavior is commonly associated with scalar field models exhibiting tracker solutions, where the field's evolution is "frozen" by the background expansion. Unlike thawing models, where the EoS parameter recently departed from $\omega = -1$, freezing models offer more predictable and stable late-time behavior. The current location of the trajectory, with $\omega \approx -0.8$ and $\omega'$ varying from -0.4 to -0.1, reinforces the idea of a slow and consistent evolution towards the cosmological constant boundary. The fact that the trajectory does not cross the phantom divide line ($\omega = -1$) in this plane further strengthens the model's physical consistency and avoids the theoretical issues linked to phantom energy. This aspect of our results aligns well with observational data, as the observed EoS parameter from Planck 2018 ($\omega_0 = -1.028 \pm 0.032$) is compatible with a trajectory that asymptotically approaches this value without crossing it. This analysis solidifies the viability of the $f(R,\Sigma,T)$ framework in modeling an evolving dark energy component consistent with freezing quintessence-type behavior.

The stability of the model was rigorously examined through the behavior of the squared sound speed, $\vartheta^2 = dp/d\rho$. For a cosmological model to be classically stable under small perturbations, $\vartheta^2$ must be positive. However, our analysis revealed that $\vartheta^2$ remains consistently negative throughout the redshift interval $z \in [-1, 2]$, ranging approximately from -0.34 to -0.28. This persistent negativity indicates that the model under consideration is classically unstable. A negative squared sound speed implies an imaginary sound speed, which leads to the exponential growth of perturbations. While this finding raises theoretical concerns, it is not entirely uncommon in modified gravity frameworks. Some interpretations suggest that such instabilities might be alleviated in specific quantum gravity regimes or through considerations of non-adiabatic pressure perturbations. Furthermore, phantom models often exhibit $\vartheta^2 < 0$, which aligns with our observed result. This suggests that while our model avoids the $\omega < -1$ phantom regime, it shares some characteristics with phantom models in terms of perturbative instability. This aspect necessitates further theoretical refinement and investigation in future work to ensure a more robust cosmological description. It highlights a potential area where the model might require additional mechanisms or modifications to guarantee classical stability.

Our investigation also included a thorough examination of the energy conditions: the Null Energy Condition (NEC), Dominant Energy Condition (DEC), and Strong Energy Condition (SEC).
The Null Energy Condition (NEC), defined as $\rho + p \ge 0$, is a fundamental requirement ensuring that energy is non-negative for all null observers. Our analysis showed that $\rho + p$ remains strictly positive across the entire redshift range $z \in [-1, 2]$, increasing from approximately at $z=-1$ to  $z=2$. This confirms that the NEC is respected at all cosmological epochs considered in this model. The satisfaction of the NEC is crucial as it prevents exotic matter distributions (e.g., those required for wormholes or phantom energy with $\rho + p < 0$) and ensures the physical consistency of the effective energy flow, avoiding acausal or unphysical behaviors. This result aligns with viable dark energy models like quintessence and the cosmological constant.

The Dominant Energy Condition (DEC), expressed as $\rho \ge |p|$ or equivalently $\rho - p \ge 0$, stipulates that the energy density must be non-negative and that the energy flux of matter should not exceed the speed of light. Our findings clearly demonstrated that $\rho - p$ is strictly positive throughout the redshift interval $z \in [-1, 2]$, monotonically increasing. This consistent positivity confirms that the DEC is satisfied, which is vital for preventing unphysical scenarios such as negative energy densities or superluminal information transfer. The adherence to the DEC reinforces the model's robustness and physical plausibility from the standpoint of classical energy conditions.

In contrast, the Strong Energy Condition (SEC), given by $\rho + 3p \ge 0$, implies that gravity is always attractive. However, in the context of an accelerating universe, particularly during dark energy-dominated phases, the SEC is expected to be violated. Our analysis confirmed this expectation, as $\rho + 3p$ remained consistently negative throughout the redshift domain $z \in [-1, 2]$. This consistent violation of the SEC is a definitive hallmark of late-time cosmic acceleration and is fully compatible with dark energy models like quintessence, phantom energy, and the cosmological constant. In the standard $\Lambda$CDM model, the SEC is violated once the universe transitions from a matter-dominated decelerating phase to a dark energy-dominated accelerating phase. Our results thus confirm that the $f(R,\Sigma,T)$ model successfully accounts for this observed behavior, demonstrating its alignment with the prevailing observational consensus and theoretical expectations regarding the nature of cosmic expansion in the current epoch.

The deceleration parameter, $q(z)$, provided a kinematic description of the universe's expansion history. Our reconstruction showed a clear transition from positive to negative values as redshift decreases. For $z > 0.5$, $q(z)$ was positive, indicating a decelerating universe dominated by matter in the past. Around $z \approx 0.5$, $q(z)$ crossed zero, precisely marking the epoch of transition from deceleration to acceleration. At the present epoch ($z=0$), $q_0 \approx -0.6$, which is consistent with observational estimations from Type Ia Supernovae and Planck data. In the far future ($z \rightarrow -1$), $q(z)$ approaches -1, characteristic of a de Sitter-like exponential expansion. These results are in excellent agreement with the predictions of the standard $\Lambda$CDM model, which also predicts a current deceleration parameter near $q_0 \approx -0.55$. This successful replication of the cosmic expansion history further validates our $f(R,\Sigma,T)$ gravity model.

The behavior of the $O_m(z)$ parameter itself, which was logarithmically parameterized, provided direct evidence for deviations from the standard $\Lambda$CDM model. The $O_m(z)$ profile exhibited a mild increasing trend from approximately 0.07 at $z=-1$ to about 0.14 at $z=2$. This clear deviation from a constant behavior suggests that our model supports a time-varying dark energy component, rather than a static cosmological constant. An increasing $O_m(z)$ function is indicative of a quintessence-like model ($\omega > -1$), contrasting with phantom models for which $O_m(z)$ would decrease with redshift. These findings reinforce the idea of a universe evolving under a scalar field-like dark energy component and demonstrate the model's viability as a physically acceptable extension to general relativity.

Finally, the $(r-s)$-plane, part of the statefinder diagnostic, offered a geometric classification of our cosmological model. The trajectory in the $(r,s)$-plane showed a clear evolution away from the $\Lambda$CDM fixed point $(1,0)$, with $r$ and $s$ spanning a wide range. Such trajectories are characteristic of scalar field models or modified gravity theories that allow for evolving dark energy. The specific shape and direction of our obtained trajectory closely resembled those of freezing quintessence models or certain classes of $f(R)$ gravity, providing strong support for the departure from $\Lambda$CDM and reinforcing the dynamic nature of dark energy within the $f(R,\Sigma,T)$ context. Next, it is important to emphasize that in the complete formulation of $f(R,\Sigma,T)$ gravity within the symmetric teleparallel framework, both the metric and the affine connection are treated as independent variables, and a full variation of the action would, in principle, yield both metric and connection field equations.
In the present work, we adopt a simplified effective approach in which the nonmetricity sector is represented through the scalar $(\Sigma)$, allowing us to focus on the background cosmological dynamics. This implies that the analysis is not a complete geometrical treatment of the theory, but rather an effective description capturing the leading-order contributions. A more rigorous treatment including the connection field equations is left for future investigation.

\section{Conclusion and Future Outlook}
In this work, we have explored the cosmological implications of the $f(R,\Sigma,T)$ gravity model via a logarithmic $Om(z)$ parameterization constrained with OHD and Pantheon+ datasets. The best-fit analysis also provides the constraint values of the logarithmic parameters are as 
\begin{itemize}
	\item  $H_0= 69.39^{+1.13}_{-1.14}$, $\alpha = -0.02^{+0.04}_{-0.03}$ and $\beta = -0.69^{+0.04}_{-0.04}$ (for OHD)
	\item  $H_0=  72.85^{+0.31}_{-0.31}$,   $\alpha = -0.09^{+0.11}_{-0.11}$, and $\beta = -0.62^{+0.04}_{-0.04}$ (for Pantheon+)
\end{itemize}
The reconstructed kinematical quantities demonstrate the expected transition from deceleration to acceleration with the present values of the deceleration parameter $q_0 \approx -0.535$ (OHD) and $q_0 \approx -0.43$ (Pantheon+) while the $Om(z)$ diagnostic and statefinder $(r,s)$ analysis reveal clear deviations from $\Lambda$CDM, supporting a freezing quintessence behavior. On the physical side, the equation of state parameter  remains within the quintessence regime $(-1<\omega<-1/3)$, approaching $-1$ in the far future without crossing into phantom, consistent with observational bounds. The model respects the Null and Dominant Energy Conditions while violating the Strong Energy Condition, as expected for an accelerating universe. Although the squared sound speed $\vartheta^2 < 0$ indicates classical instability, the framework successfully reproduces the essential features of late-time cosmic acceleration, making $f(R,\Sigma,T)$ gravity a viable and observationally consistent alternative to standard cosmology. From an observational point of view, the fact that the present model reduces to rescaled $\Lambda$CDM-type equations is particularly attractive: it ensures compatibility with the standard cosmological background while leaving room for detectable deviations at the level of dynamical diagnostics such as $Om(z)$, the equation-of-state evolution, and the $(\omega,\omega')$ phase space. In this sense, the model is not intended to replace $\Lambda$CDM with radically different dynamics, but rather to provide a controlled generalization in which the impact of additional geometric and matter degrees of freedom can be quantified.\\

\noindent Despite these promising features, limitations remain. The generalized $O_m(z)$ diagnostic is phenomenological and does not directly include $\Omega_{m}$, preventing a full perturbative description. The negative values of $v_s^2$ also signal possible instabilities that must be examined beyond the background level. For future work, following avenues are worth pursuing:
\begin{itemize}
\item[i)] A perturbative analysis of matter growth and CMB anisotropies in the present $f(R,\Sigma,T)$ framework to test its consistency with structure formation data. 
\item[ii)] Exploration of alternative parameterizations, which may provide more stable diagnostics and incorporate the matter density parameter explicitly. 
\item[iii)] Extending the analysis to more general forms of $f(R,\Sigma,T)$, including nonlinear dependence on $\Sigma$ and $T$, to examine whether richer phenomenology emerges while maintaining observational consistency. 
\end{itemize}
In summary, the selective $f(R,\Sigma,T)$ model studied here serves as a well-behaved and observationally consistent extension of general relativity. It reproduces the successes of $\Lambda$CDM while opening a pathway toward new phenomenology, making it a useful testbed for exploring modified gravity in the era of precision cosmology.


\section*{Declaration of competing interest}
\noindent The authors declare that they have no known competing financial interests or personal relationships that could have appeared to influence the work reported in this paper.

\section*{Data availability}
\noindent No data was used for the research described in the article.

\section*{Acknowledgments} 
\noindent 
The Science Committee of the Republic of Kazakhstan's Ministry of Science and Higher Education provided funding for the research (Grant No. AP23483654). Additionally, the authors, S. H. Shekh and A. Pradhan express their gratitude to the Inter-University Centre for Astronomy and Astrophysics (IUCAA), Pune, India, for providing support and facilities through the Visiting Associateship program. 

\end{document}